\documentclass[aps,amsmath,amssymb,floatfix,lengthcheck,preprintnumbers,prd,showpacs,superscriptaddress,twocolumn]{revtex4-1} 

\usepackage{hyperref}
\usepackage{verbatim}
\usepackage{graphicx}
\usepackage{color}
\usepackage{bm} % bold math
\usepackage{ifpdf} % This lets us get figures right both in pdf and ps versions

\newcommand{\NRQCD}{\textrm{NRQCD}}
\newcommand{\NR}{\textrm{NR}}
\renewcommand{\vec}[1]{{\bm{#1}}}
\newcommand{\bs}{\mbox{\boldmath $\sigma$}}
\newcommand{\bA}{\mbox{\boldmath $A$}}

\newcommand{\bk}{\mbox{\boldmath $k$}}
\newcommand{\boldD}{\mathbf{D}}
\newcommand{\boldE}{\mathbf{E}}
\newcommand{\boldB}{\mathbf{B}}
\newcommand{\boldsigma}{{\bm\sigma}}
\newcommand{\boldDelta}{{\bm\Delta}}
\newcommand{\BE}{\begin{equation}}
\newcommand{\EE}{\end{equation}}
\newcommand{\BEA}{\begin{eqnarray}}
\newcommand{\EEA}{\end{eqnarray}}
\newcommand{\BES}{\begin{eqnarray*}}
\newcommand{\EES}{\end{eqnarray*}}
\newcommand{\nn}{\nonumber}
\newcommand{\be}{\begin{equation}}
\newcommand{\ee}{\end{equation}}
\def\bea{\begin{align}}
\def\ena{\end{align}}
\def\beqa{\begin{eqnarray}}
\def\enqa{\end{eqnarray}}
\newcommand{\gpartial}{\displaystyle{\not} \partial}

\newcommand{\gA}{\displaystyle{\not} A}
\newcommand{\Psib}{\bar{\Psi}}
\newcommand{\psib}{\bar{\psi}}
\newcommand{\ra}{\rangle}
\newcommand{\la}{\langle}

\newcommand{\bacol}{\setlength{\arraycolsep}{0pt}}

\newcommand{\cambridge}{Department~of~Applied~Mathematics~and~Theoretical~Physics, University~of~Cambridge, Centre~for~Mathematical~Sciences, Cambridge~CB3~0WA, United~Kingdom}
\newcommand{\cray}{Cray~Exascale~Computing~Research~Initiative, University~of~Edinburgh, King's~Buildings, Mayfield~Road, Edinburgh~EH9~3JZ, United~Kingdom}
\newcommand{\edinburgh}{SUPA, School~of~Physics~and~Astronomy, University~of~Edinburgh, King's~Buildings, Mayfield~Road, Edinburgh~EH9~3JZ, United~Kingdom}
\newcommand{\mainz}{Institut~f\"ur~Kernphysik, University~of~Mainz, Becher-Weg~45, 55099~Mainz, Germany}

\begin{document}

\title{Radiative improvement of the lattice NRQCD action using the background 
field method with applications to quarkonium spectroscopy}

\author{T.C. \surname{Hammant}}
\affiliation{\cambridge}
\author{A.G. \surname{Hart}}
\affiliation{\cray}\affiliation{\edinburgh}
\author{G.M. \surname{von Hippel}}
\affiliation{\mainz}
\author{R.R. \surname{Horgan}}
\author{C.J. \surname{Monahan}}
\affiliation{\cambridge}

\pacs{12.38.Bx, 12.38.Gc}
\preprint{DAMTP-2013-14}
%\preprint{MKPH-T-11-....}

\begin{abstract} 

We apply the background field~(BF) method to Non-Relativistic QCD (NRQCD) on the 
lattice in order to determine the one-loop radiative corrections to the coefficients 
of the NRQCD action in a manifestly gauge-covariant manner by matching the NRQCD 
prediction for particular on-shell processes with those of relativistic continuum 
QCD. We explain how the BF method is implemented in automated perturbation theory and 
discuss the technique for matching the relativistic and non-relativistic theories. We 
compute the one-loop radiative corrections to the $\vec{\sigma}\cdot\vec{B}$ and 
Darwin terms for the NRQCD action currently used in simulations, as well as the 
one-loop coefficients of the spin-dependent $O(\alpha^2)$ four-fermion contact terms. 
The effect of the corrections on the hyperfine splitting of bottomonium is estimated 
using earlier simulation results
\cite{PhysRevD.72.094507};
the corrected lattice prediction is found to be in agreement with experiment.
Agreement of the hyperfine splitting of bottomonium and the $B$-meson system is confirmed by 
recent simulation studies 
\cite{Dowdall:2011wh,Dowdall:2012ab}
which include our NRQCD radiative corrections for the first time.

\end{abstract}

\maketitle

%%%%%%%%%%%%%%%%%%%%%%%%%%%%%%%%%%%%%%%%%%%%%%%%%%%%%%%%%%%%%%%%%%%%%%%%%%%%%%%%
%%%%%%%%%%%%%%%%%%%%%%%%%%%%%%%%%%%%%%%%%%%%%%%%%%%%%%%%%%%%%%%%%%%%%%%%%%%%%%%%
% Include contents
%%%%%%%%%%%%%%%%%%%%%%%%%%%%%%%%%%%%%%%%%%%%%%%%%%%%%%%%%%%%%%%%%%%%%%%%%%%%%%%%
%%%%%%%%%%%%%%%%%%%%%%%%%%%%%%%%%%%%%%%%%%%%%%%%%%%%%%%%%%%%%%%%%%%%%%%%%%%%%%%%
%%%%%%%%%%%%%%%%%%%%%%%%%%%%%%%%%%%%%%%%%%%%%%%%%%%%%%%%%%%%%%%%%%%%%%%%%%%%%%%%
%%%%%%%%%%%%%%%%%%%%%%%%%%%%%%%%%%%%%%%%%%%%%%%%%%%%%%%%%%%%%%%%%%%%%%%%%%%%%%%%
\section{Introduction}
%%%%%%%%%%%%%%%%%%%%%%%%%%%%%%%%%%%%%%%%%%%%%%%%%%%%%%%%%%%%%%%%%%%%%%%%%%%%%%%%
%%%%%%%%%%%%%%%%%%%%%%%%%%%%%%%%%%%%%%%%%%%%%%%%%%%%%%%%%%%%%%%%%%%%%%%%%%%%%%%%
\label{sec:intro}

The decays of hadrons containing bottom quarks provide some of the most
stringent tests of the Standard Model~(SM) so far, but direct simulations
of the bottom quark in lattice QCD remain fraught with problems due to
discretization effects arising from the large mass of the bottom quark in
conjunction with current limitations on achievable lattice spacings.
An alternative is provided by Non-Relativistic QCD~(NRQCD)
\cite{PhysRevD.46.4052},
an effective field theory for heavy quarks whose use in describing
heavy-flavour hadrons has so far met with considerable success
\cite{PhysRevD.72.094507}
and which represents one important approach to obtaining accurate predictions
for flavour physics observables and testing the limits of the SM.
However, until recently the NRQCD actions used in lattice simulations did
not include radiative corrections to the action, limiting the accuracy
to which such important quantities as the $\Upsilon$--$\eta_{\rm b}$ hyperfine
splitting could be predicted
\cite{Meinel10}.
This is in stark contrast to the situation of Non-Relativistic QED~(NRQED),
for which the radiative improvements to the action have long been known,
leading to highly precise theoretical predictions for muonium hyperfine
structure and for positronium decay
\cite{Kinoshita:1995mt,PhysRevD.55.7267,Labelle:1997}.
Achieving similar precision for NRQCD predictions by including the radiative
improvements in the NRQCD action is therefore highly desirable. However,
a crucial difference to the NRQED case is that the strongly interacting
non-abelian nature of QCD and NRQCD imposes confinement and calls for a
lattice implementation of NRQCD; this means that the complete $1/(Ma)^n$
structure of all quantities must be retained, as opposed to the situation
in (continuum) NRQED, where there are ways to drop terms in $(\Lambda/M)^n$
consistently. A further complication arises from infrared~(IR) divergences,
which turn out to play a significant r\^ole in the non-abelian case.

In this paper, we follow up on our letter
\cite{PhysRevLett.107.112002}
where we presented the first calculation of radiative corrections to the
lattice NRQCD action using the background field~(BF) method.
We proceed by computing the effective action in both continuum QCD and
lattice NRQCD at one loop and matching the latter term by term to
the non-relativistic reduction of the former.
In particular, we determine the $O(\alpha_s)$ correction
to the coefficient $c_4$ of the chromomagnetic $\vec{\sigma}\cdot\vec{B}$
operator and the leading contributions to the coefficients of the spin-dependent four-fermion
contact operators in the NRQCD action, as required in order to enable a more
precise determination of the hyperfine structure of heavy quarkonia in NRQCD.
Using our results, we are able to estimate the $O(\alpha_s^2)$ correction
to the hyperfine splitting of the S-wave bottomonium states using
simulation data from
\cite{PhysRevD.72.094507},
giving a corrected value of $72(3)(5)(3)$~MeV, which agrees with the
experimental value of $69.3(2.8)$~MeV
\cite{PDG:2010}.%
The correction has the additional beneficial effect of reducing the lattice
spacing dependence, placing the remaining $O(a^2)$ discretization errors well
below other sources of error.

We remark that our calculation reported here completes the one-loop radiative improvement
of quark bilinear terms to ${\cal O}(v^4)$ in the NRQCD action for which the improvement 
of the purely kinetic terms was reported in \cite{Dowdall:2011wh}: see section \ref{sec:NRQCDaction}. 

In section \ref{sec:BFNRQCD} we discuss in some more detail the approach
to matching using the BF method  and discuss the gauge invariance of our
results;
in sections \ref{sec:chromomag} and \ref{sec:darwin}, we proceed to
apply the BF method to the matching of the chromomagnetic and Darwin
terms in the NRQCD action.
The four-fermion interactions in NRQCD are the topic of section \ref{sec:4f},
where we derive results for the spin-dependent four-fermion operators
and explain the difficulties in matching their spin-independent counterparts,
which will be the topic of a future paper.
We apply our results to the hyperfine splitting and S-wave mass shifts
in section \ref{sec:application}
before summarizing our results in section \ref{sec:conclusions}.

In the following we denote the perturbative expansion for a generic quantity $w$
as $w = \sum_{n=0}w^{(n)}\alpha_s^n$. Where $w^{(n)}$ contains IR divergences,
the IR finite part is denoted with a tilde: $\widetilde{w}^{(n)}$.

%%%%%%%%%%%%%%%%%%%%%%%%%%%%%%%%%%%%%%%%%%%%%%%%%%%%%%%%%%%%%%%%%%%%%%%%%%%%%%%%
%%%%%%%%%%%%%%%%%%%%%%%%%%%%%%%%%%%%%%%%%%%%%%%%%%%%%%%%%%%%%%%%%%%%%%%%%%%%%%%%
\section{The Background field method for lattice NRQCD}
%%%%%%%%%%%%%%%%%%%%%%%%%%%%%%%%%%%%%%%%%%%%%%%%%%%%%%%%%%%%%%%%%%%%%%%%%%%%%%%%
%%%%%%%%%%%%%%%%%%%%%%%%%%%%%%%%%%%%%%%%%%%%%%%%%%%%%%%%%%%%%%%%%%%%%%%%%%%%%%%%
\label{sec:BFNRQCD}

The BF method
\cite{DeWitt:1967ub,DeWitt:1967uc,KlubergStern:1974xv,Abbott:1983,Vilkovisky:1984}
is well established as a tool to compute the effective action in quantum
field theory, which has a number of very attractive features: firstly,
QCD in background field gauge (BFG) satisfies a set of abelian-like Ward
identities reducing the number of calculations needed in order to renormalize
it. Secondly, the existence of a residual gauge invariance in BFG implies
that only gauge-covariant operators can appear in the effective action, which
is of particular importance when considering operators of dimension $D>4$
as required in an effective theory, where a loss of gauge-covariance would
herald a proliferation of additional operators.

A method to derive a unique effective action is given by the Vilkovisky-DeWitt~(VDW) 
technique and one choice is to use the Landau-DeWitt gauge corresponding to the BF 
generalization of Landau gauge
\cite{Rebhan1987832}.
Whether or not this approach is applicable to our matching procedure remains to be 
investigated. However, for our purposes, the VDW method is not necessary since we are 
able to perform our matching solely using on-shell quantities including S-matrix 
elements which can be given an unambiguous physical interpretation.

The BF method is indispensable to our matching procedure. As an effective theory, 
NRQCD has operators of dimension $D>4$ appearing in the action, and BFG is needed to 
ensure a gauge-covariant form of the effective action; this is not guaranteed nor 
likely without using the BF formalism. It should be noted that the appearance of 
gauge-non-covariant $D>4$ operators in the effective action would not in and of 
itself be incorrect; however, by obscuring the gauge symmetry of the theory, they 
would hinder a physical interpretation of the results and significantly complicate 
the calculations. Moreover, without the use of BFG, a further source of complexity 
would arise from the appearance of ultraviolet~(UV) logarithms in the coefficients of 
operators in the effective action, whose contributions to physical processes would 
have to cancel against the contributions from the additional non-gauge-covariant 
operators.

Finally, the BF method makes our work easier and tractable because the QED-like Ward 
identities in BFG are enough to render the one-particle irreducible~(1PI) vertex 
function finite, so that only the gluonic self-energy renormalizes the coupling, 
whereas the BF itself is not renormalized. Since these statements hold in both QCD and 
NRQCD, we are able to match the two theories by equating two quantities that are UV 
finite and thus we can use different regulators on either side. In particular, the 
QCD vertex can be calculated analytically in the continuum using dimensional 
regularization~(DR), or numerically on fine lattices approaching the continuum limit, 
where the latter is particularly convenient for checking the gauge-parameter 
independence of the result since analytical calculations for arbitrary values of the 
gauge parameter tend to become rather involved.

The remaining issue is how to regularize the infrared (IR) divergences which arise at 
intermediate steps in our calculation. The radiative corrections to the coefficients 
in the NRQCD action are, and must be, IR finite: they cannot depend on any scheme to 
regularize IR divergences. Two kinds of IR divergences arise. The first are those 
that occur in the continuum calculation of the diagrams. These divergences are 
a-priori independent of the non-zero lattice spacing $a$ and must match directly 
between the relativistic QCD and NRQCD calculations. The second are lattice artifact 
IR divergences that occur solely in the NRQCD one-loop calculations, and the 
contributions from the set of diagrams under consideration must cancel consistently 
within the NRQCD calculation since they depend on $a$. That the IR divergences match 
or cancel in this way is a strong consistency check on our calculations. In NRQCD we 
are also able to calculate both kinds of IR divergence analytically in every case 
considered and also numerically and agreement between the two approaches is a further 
check on our results.

Whilst it might, in principle, be possible to eliminate the continuum IR divergences 
by subtracting the integrand in continuum QCD from the equivalent one in NRQCD, this 
approach is numerically very difficult. It is compounded by the need to also 
eliminate the lattice artifact IR divergences in the same way, otherwise the use of 
an IR regulation scheme is, in any case, inevitable; this would be numerically very 
hard to accomplish. We cannot use dimensional regularization to regulate the IR 
divergences since these are not applicable for lattice field theories. There are a 
number of other methods that can be used in our case. 

Twisted boundary conditions (TWB) can be applied corresponding to matching 
the non-abelian gauge theories in a region of finite extent $L$ in two or three 
spatial dimensions \cite{'tHooft:1979uj,Weisz84,Luscher1986309}. The TWB eliminate 
the zero modes for the non-abelian gauge field and give a gauge invariant regulation 
of IR divergences with a mass-scale $2\pi/3L$ for the $SU(3)$ gauge theory. Although 
there are extra complications in the background field approach, TWB are relatively 
straightforward to implement for numerical lattice calculations and have been used to 
carry out improvement calculations \cite{Luscher1986309,Hart:2008sq}. However, they 
are difficult to implement analytically for the continuum QCD calculations and 
although there is progress in this area we choose not to implement the TWB.

The NRQCD action is a derivative expansion and the strategy is to match 
the expansion in independent external momenta for each amplitude considered. IR 
divergences can be regulated by using a small but non-zero momentum transfer, $q$ 
say, in the diagram concerned. After cancellation of the IR singularities we may 
safely set $q=0$. The application of this method to diagrams with more than one loop 
needs careful thought but is certainly an option for one-loop calculations.

We regulate the IR divergences in both QCD and NRQCD using a gluon mass 
$\mu$. This is known to be correct for diagrams which have a QED-like topology: in 
QCD the difference is a simple overall colour factor. In general, a non-zero gluon 
mass in QCD breaks gauge invariance and in our calculation the one-loop diagrams for 
matching the quark bi-linear operators, which give the chromodynamic form factors, contain 
three-gluon vertices and the use of a gluon mass IR regulator needs justification. We 
note that at one-loop the diagrams are identical to those that would arise in the Curci-Ferrari 
theory \cite{Curci:1976bt, Curci:1976kh} which is renormalizable and known to recover 
QCD in the $\mu \to 0$ for gauge-invariant quantities; the introduction of a non-zero gluon mass by extending 
QCD to the Curci-Ferrari formulation is a mechanism for regulating IR divergences 
\cite{Curci:1976bt,Curci:1976kh,Curci:1978ku,Ojima:1981fs, 
Blasi:1995vt,Gracey:2001ma,Gracey:2002yt,Browne:2002wd, Kondo:2012ri}. We match on-shell 
gauge-invariant physical processes and in our final 
result all IR divergences cancel to give IR-finite gauge-covariant counterterms in NRQCD
and the limit $\mu \to 0$ can be taken. Diagrams with more than one loop in Curci-Ferrari 
theory will contain extra contributions compared with the same process in QCD. The 
Curci-Ferrari theory on a lattice is formulated and discussed by von Smekal et al. 
\cite{vonSmekal:2008en}.

For the case of matching the quark bi-linear operators another approach is to 
introduce gluon masses by the spontaneous symmetry breaking of $SU(3)\to U(1)_3 
\times U(1)_8$ \cite{Peskin:1995ev} and calculate in the renormalizable $R_\xi$ 
gauge. The unbroken abelian gauge groups are generated, respectively, by the $T_3$ 
and $T_8$ $SU(3)$ generators. In general, QCD is not recovered in the zero mass limit 
since there remain massless scalar fields. However, the one-loop diagrams that we 
calculate are identical in both QCD and the spontaneously broken theory. In this 
case, our present calculation corresponds to computing the quark bi-linear form factors for the 
$U(1)_8$ (massless) gluon.

An alternative justification was established by Kautsky \cite{Kautsky:2006ab}. In BFG the Ward 
Identities in QCD give a QED-like relationship between the quark wavefunction 
renormalization constant, $Z_2$, and the vertex renormalization constant, $Z_1$, 
namely $Z_1=Z_2$; a violation of this equality would signal a breakdown of gauge 
invariance. Kautsky showed in BFG that at one loop this equality does hold when the 
IR divergences in the diagrams concerned are regulated by a non-zero gluon mass. This 
is a crucial test of this approach for regulating IR divergences.

An important observation, that we verify, is that the results of calculating the physical process 
used for the matching procedure in NRQCD and QCD are separately independent of the gauge
parameter multiplying the BF gauge-fixing term in the action: the NRQCD radiative counterterms are
both gauge-covariant and independent of the gauge parameter.

%%%%%%%%%%%%%%%%%%%%%%%%%%%%%%%%%%%%%%%%%%%%%%%%%%%%%%%%%%%%%%%%%%%%%%%%%%%%%%%%
%%%%%%%%%%%%%%%%%%%%%%%%%%%%%%%%%%%%%%%%%%%%%%%%%%%%%%%%%%%%%%%%%%%%%%%%%%%%%%%%
\section{The NRQCD action}
%%%%%%%%%%%%%%%%%%%%%%%%%%%%%%%%%%%%%%%%%%%%%%%%%%%%%%%%%%%%%%%%%%%%%%%%%%%%%%%%
%%%%%%%%%%%%%%%%%%%%%%%%%%%%%%%%%%%%%%%%%%%%%%%%%%%%%%%%%%%%%%%%%%%%%%%%%%%%%%%%
\label{sec:NRQCDaction}

The lattice NRQCD action up to and including $O(v^6)$ operators is given by
\cite{PhysRevD.46.4052}
\begin{equation}
S_{NRQCD}=\sum_{x,\tau}\psi^\dagger(x,\tau)\left[\psi(x,\tau)-K(\tau)\psi(x,\tau-a)\right]\; ,
\end{equation}
with the kernel
\begin{widetext}
\begin{equation}
K(\tau)=\left( 1-\frac{a\delta H}{2}\right) \left( 1-\frac{aH_0}{2n}\right) ^{n}U^\dagger_4
\left( 1-\frac{aH_0}{2n}\right) ^{n}\left( 1-\frac{a\delta H}{2}\right)\; ,
\end{equation}
\end{widetext}
where $H_0$ is the standard NRQCD kinetic operator, $H_0=\Delta^{(2)}/2M$.

The quark Green function satisfies the evolution equation
\begin{equation}
G(\mathbf{x},t+a)=K(t+a)G(\mathbf{x},t)+\delta_{x,0}\delta_{t+a,0}\; ,
\label{eq:NRQCDgreen}
\end{equation}
where $G(\mathbf{x},t)$ vanishes for $t<0$. The kernel $K$ is constructed
so as to give an evolution that is symmetric with respect to time reversal,
and leads to a smaller wave function renormalization than some other
formulations
\cite{Davies:1991py}.
The parameter $n$ was introduced to prevent instabilities at large momenta due
to the kinetic energy operator
\cite{PhysRevD.43.196}
(it is easy to see that the Green function defined by Eq. (\ref{eq:NRQCDgreen})
diverges if ${aH_0/2n}>2$). For a given $\beta$ and lattice spacing we
therefore have a minimum $n$ value. For the values of lattice spacing used by 
the HPQCD collaboration $n=2$ or $n=4$ suffices.

The interaction terms for the ``full NRQCD'' action are given by,
\begin{eqnarray}
\delta H_{full}&=&-c_1\frac{(\Delta^{(2)})^2}{8(Ma)^3}\nn\\
&+&c_2\frac{ig}{8M^2}\left({\boldDelta}^{(\pm)}\cdot\widetilde{\boldE}-\widetilde{\boldE}
\cdot {\boldDelta}^{(\pm)}\right)\nn\\
&-&c_4\frac{g}{2M}\boldsigma\cdot\widetilde{\boldB}-c_3\frac{g}{8M^2}\boldsigma\cdot\left(\widetilde{\boldDelta}^{(\pm)}\times\widetilde{\boldE}-\widetilde{\boldE}\times\widetilde{\boldDelta}^{(\pm)}\right)\nonumber\\
&+&c_5\frac{\Delta^{(4)}}{24M}-c_6\frac{(\Delta^{(2)})^2}{16nM^2}\; ,
\label{eq:v4}
\end{eqnarray}
including operators with coefficients $c_5$ and $c_6$ to remove the leading
discretization artifacts in the improved NRQCD kinetic operator
at $O(a^4p^4)$. The operators have been normalized such that $c_i=1$
at tree level. The one-loop radiative corrections to $c_1, c_5$ and $c_6$ have been reported in \cite{Dowdall:2011wh}.
We can also add further spin-dependent interaction terms to obtain the ``full spin $v^6$ NRQCD'' action given by
\begin{eqnarray}
\delta H_{v^6}&=&-f_1\frac{g}{8M^3}\left\lbrace {\Delta^{(2)},{\boldsigma}\cdot\widetilde{\boldB}}\right\rbrace \nn\\
&&-f_2\frac{3g}{64M^4}\left\lbrace {\Delta^{(2)},{\boldsigma}\cdot
\left(\widetilde{\boldDelta}^{(\pm)}\times\widetilde{\boldE}-\widetilde{\boldE}\times\widetilde{\boldDelta}^{(\pm)}\right)}\right\rbrace \nn\\
&&-f_3\frac{ig^2}{8M^3}{\boldsigma}\cdot\widetilde{\boldE}\times\widetilde{\boldE}\; ,
\label{eq:v6}
\end{eqnarray}
where again the operators have been normalized such that at tree-level $f_i=1$.

The unimproved forward and backward derivatives are defined by
\begin{eqnarray}
(\Delta^{+}_\mu f)(x)&=&\frac{1}{a}(U_\mu(x)f(x+\hat{\mu})-f(x)), \nonumber\\
(\Delta^{-}_\mu f)(x)&=&\frac{1}{a}(f(x)-U_{-\mu}(x)f(x-\hat{\mu}))\; ,
\label{eq:fwdbckderiv}
\end{eqnarray} 
and the unimproved higher order lattice derivatives are
\begin{equation}
\Delta^{(2n)}=\sum_{j=1}^{3}(\Delta^{(+)}_j\Delta^{(-)}_j)^n\; . 
\end{equation}
The $O(a^4p^4)$-improved symmetric derivative is given by
\begin{equation}
\widetilde{\Delta}^{(\pm)}=\Delta_j^{(\pm)}-\frac{1}{6}\Delta^{(+)}_j\Delta^{(\pm)}_j\Delta^{(-)}_j\; .
\end{equation}

$\widetilde{\boldB}$ and $\widetilde{\boldE}$ are the improved chromomagnetic
and chromoelectric fields, constructed from standard cloverleaf operators,
and improved to $O(a^2)$:
\begin{align}
&\widetilde{B_k}=-\frac{1}{2}\epsilon_{klm}F^{imp}_{lm},\\ 
&\widetilde{E_k}=F^{imp}_{k4} \label{eq:imprB},\\
&F^{imp}_{\mu\nu}=\frac{5}{3}F_{\mu\nu}-\frac{1}{6}[U_\mu F_{\mu\nu} U^\dagger_\nu+U^\dagger_\mu F_{\mu\nu} 
U_\nu + (\mu \leftrightarrow \nu) ],\\
&F_{\mu\nu}=-\frac{i}{2g}(I_{\mu\nu}-I_{\mu\nu}^\dagger),\\
&I_{\mu\nu}= \nonumber\\
&\frac{1}{4}\sum_{\{(\alpha,\beta)\}_{\mu\nu}}U_\alpha(x)U_\beta(x+\hat{\alpha})
                  U_{-\alpha}(x+\hat{\alpha}+\hat{\beta})U_{-\beta}(x+\hat{\beta})\; ,
\end{align}
where
$\{(\alpha,\beta)\}_{\mu\nu}={(\mu,\nu),(\nu,-\mu),(-\mu,-\nu),(-\nu,\mu)}$
for $\mu\ne\nu$. 

Mean-field (tadpole) improvement is implicitly included in the definition of the 
links by replacing $U_\mu\mapsto U_\mu/(u_0)$, where $u_0$ is the Landau gauge
mean link. Since the gauge links are unitary, some of the correction terms
in $F^{imp}$ will be only four, rather than six, links long due to
cancellations of the form $U_\mu U_\mu^\dag$, and so should carry a factor
of $\frac{1}{u_0^4}$ (rather than $\frac{1}{u_0^6}$). This is implemented
by the addition of a further correction term
\cite{Groote:2000jd}
and must be taken into account when calculating mean-field improvement
contributions.

%%%%%%%%%%%%%%%%%%%%%%%%%%%%%%%%%%%%%%%%%%%%%%%%%%%%%%%%%%%%%%%%%%%%%%%%%%%%%%%%
%%%%%%%%%%%%%%%%%%%%%%%%%%%%%%%%%%%%%%%%%%%%%%%%%%%%%%%%%%%%%%%%%%%%%%%%%%%%%%%%
\section{Chromomagnetic Interaction}
%%%%%%%%%%%%%%%%%%%%%%%%%%%%%%%%%%%%%%%%%%%%%%%%%%%%%%%%%%%%%%%%%%%%%%%%%%%%%%%%
%%%%%%%%%%%%%%%%%%%%%%%%%%%%%%%%%%%%%%%%%%%%%%%%%%%%%%%%%%%%%%%%%%%%%%%%%%%%%%%%
\label{sec:chromomag}

The hyperfine splitting of S-wave bottomonium, $M_\Upsilon-M_{\eta_b}$,
provides a high-precision test of NRQCD
\cite{Meinel10}.
The size of the hyperfine splitting is expected to be
approximately proportional to the square of the coefficient $c_4$
of the chromomagnetic operator in the NRQCD action.
While most NRQCD simulations so far have set $c_4=1$, this
coefficient will receive radiative corrections at $O(\alpha_s)$,
which have the potential of affecting the predicted value for the
hyperfine splitting significantly, so that without a determination
of these radiative corrections, large systematic errors have to be
included in any NRQCD calculation of the hyperfine splitting.
For example, in
\cite{PhysRevD.72.094507}
the ground state hyperfine splitting is predicted to be
$M_{\Upsilon}-M_{\eta_b}=61(14)MeV$, where the dominant errors are
from the missing $O(\alpha_s)$ corrections to $c_4$.
This prediction is smaller than the experimental value of
$M_{\Upsilon}-M_{\eta_b}=71.4(+2.3)(-3.1)(3.7)$,
\cite{PhysRevLett.101.071801} 
although within the large theoretical errors, theory and experiment agree.

\begin{figure}
\centering
\includegraphics[width=0.85\linewidth,clip=true]{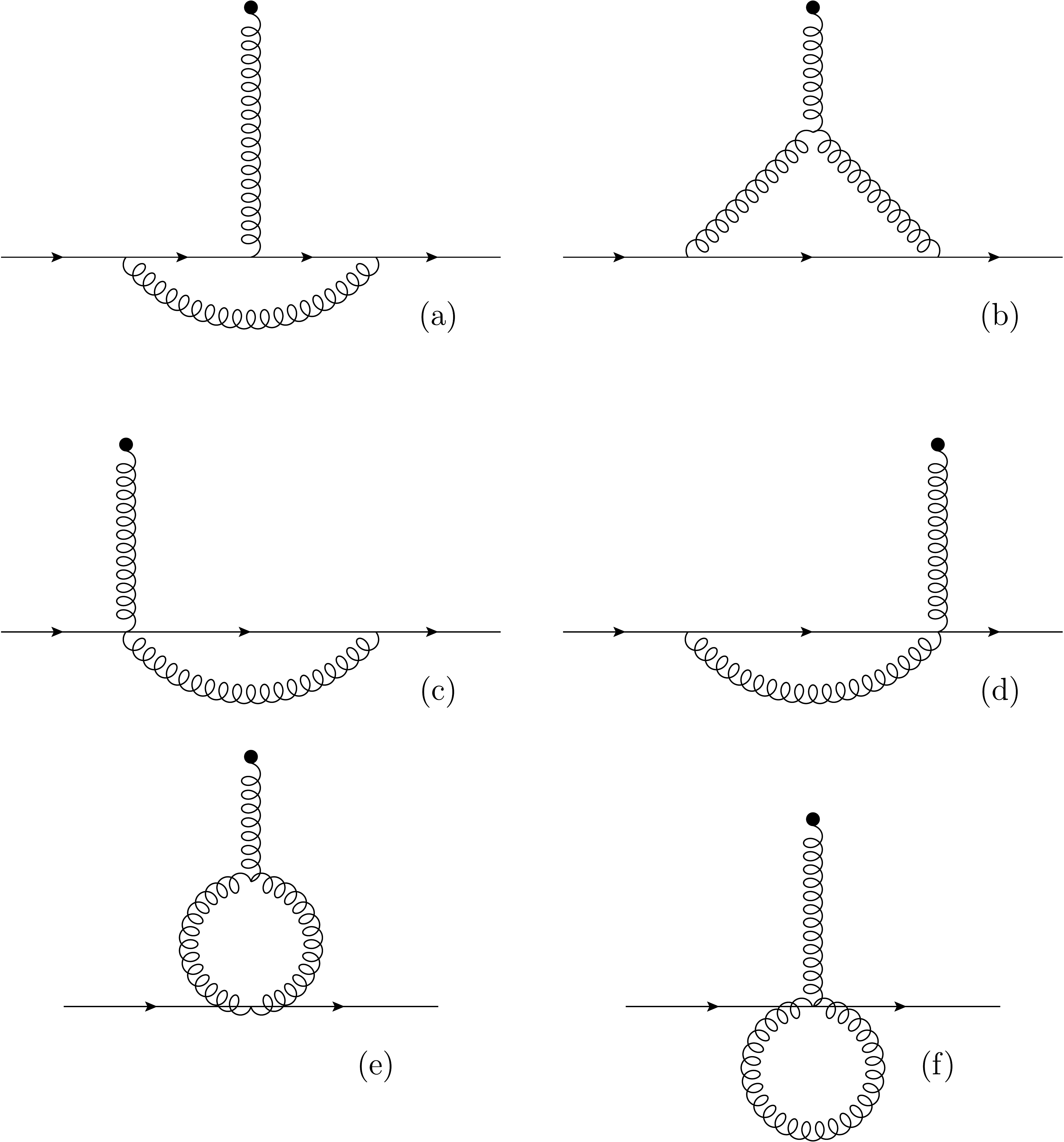}\;
\caption{Vertex correction Feynman diagrams.
         Referred to as (a) QED (b) nonabelian (c)\&(d) swordfish (e) algae
         (f) ankh diagrams.}
\label{fig:vertexdiags}
\end{figure}

At the one-loop level, the contributions to the effective action
relevant for determining the chromomagnetic operator arise from the
vertex diagrams shown in Figure \ref{fig:vertexdiags}.
As the QCD analogue of the magnetic moment operator in QED,
the chromomagnetic moment operator constitutes the leading
contribution to the hyperfine splitting within hadronic states,
and by tuning of the coefficient $c_4$ of the chromomagnetic operator
in the NRQCD action, we can match NRQCD to QCD as far as the hyperfine
splitting is concerned.

In this section we compute the radiative improvement correction
to $c_4$ which takes the form $c_4 = 1 + \alpha_s c_4^{(1)}$
with $c_4^{(1)} = A + B \log(Ma)$,
where $a$ is the lattice spacing, $M$ is the heavy quark mass,
and $A$ and $B$ are constants which we calculate. 
In continuum QCD, all corrections can be computed analytically
following standard techniques; for the calculation in lattice NRQCD,
we employ the automated Feynman rule packages
\textsc{HPsrc} and \textsc{HiPPy}
\cite{Hart20092698}.

%%%%%%%%%%%%%%%%%%%%%%%%%%%%%%%%%%%%%%%%%%%%%%%%%%%%%%%%%%%%%%%%%%%%%%%%%%%%%%
\subsection{Continuum QCD Calculation}
\label{subsec:chromocont}

The QCD effective action contains the following terms bi-linear in the
fermionic fields,
\begin{equation}
 \Gamma^{QCD}= Z_{2}^{-1}\Psib(\gpartial+\gA)\Psi+\delta Z_\sigma \Psib\frac{\sigma^{\mu\nu}F_{\mu\nu}}{2m}
\Psi+\ldots\ , 
\end{equation}
where, because of BFG invariance, the matter terms and $\Psib\gA\Psi$ vertex
are packaged into a gauge covariant derivative so as to obey the
Slavnov-Taylor identity $Z_{1F}=Z_2^{-1}$. The constants $Z$ are thus
related to the usual chromodynamic form factors via $Z_{1F}=F_1(0)$
and $\delta Z_\sigma = F_2(0)$. 
The first term is multiplicatively renormalized to
\begin{equation}
 \Psib_R(\gpartial+\gA)\Psi_R,
\label{eq:gArenorm}
\end{equation}
and BFG ensures that $A$ is not renormalized. The second term renormalizes to
\begin{equation}
 b_\sigma\Psib_R\frac{\sigma^{\mu\nu}F_{\mu\nu}}{2m_R}\Psi_R,
\label{eq:sigmarenorm}
\end{equation}
with
\begin{equation}
b_\sigma=Z_\sigma Z_2 Z_m.
\end{equation}
Since QCD is renormalizable, the absence of a Pauli term in the bare
QCD action ensures that $b_\sigma$ is UV finite; it moreover implies
that $Z_\sigma=O(\alpha_s)$, so that to leading order we can set
$Z_2 = Z_m = 1$.

After performing a non-relativistic reduction using a
Foldy-Wouthuysen-Tani~(FWT) transformation
\cite{PhysRev.78.29,PTP.6.267}, 
Eqs. (\ref{eq:gArenorm}) and (\ref{eq:sigmarenorm}) reduce to
\begin{equation}
(1+b_\sigma)g\psi_R^\dagger\frac{\boldsigma\cdot\boldB}{2m_R}\psi_R,
\label{eq:NRred}
\end{equation}
where $B$ is the chromomagnetic field.
It is worth noting the relationship between various notations
for these renormalization constants: at the one loop level
\begin{equation}
1+b_\sigma=1+Z_\sigma Z_2=(F_1(0)+F_2(0))Z_2 = G_M(0)Z_2\;. 
\end{equation}
Note that there is no factor of $Z_m$ in these expressions,
since the Gordon decomposition of the tree-level $\Psib\gA\Psi$ 
term is between on-shell spinors and so we automatically have the
renormalized mass in the resulting chromomagnetic term.
As stated before, $b_\sigma$ is UV finite; this is crucial for our
analysis, since it enables us to directly equate results obtained on
the lattice to those obtained in the continuum since the difference
between the schemes for UV regulation is then irrelevant.

The two diagrams contributing to $b_\sigma$ are given in Figure
\ref{fig:vertexdiags} (a) and (b), and we
introduce a small gluon mass $\mu$ as an IR regulator.
A straightforward analytic calculation then gives
\begin{equation}
b_\sigma=\frac{3\alpha_s}{2\pi}\log{\frac{\mu}{m}}+\frac{13\alpha_s}{6\pi},
\end{equation}
at one loop. This continuum calculation was verified using the
\textsc{HiPPy} and \textsc{HPsrc} packages by reproducing the same result
numerically in the limit $ma\rightarrow0$; in this way it was also confirmed
that the result was gauge-parameter independent.
Note that the choice of BFG was crucial to ensure a UV finite result.

%%%%%%%%%%%%%%%%%%%%%%%%%%%%%%%%%%%%%%%%%%%%%%%%%%%%%%%%%%%%%%%%%%%%%%%%%%%%%%
\subsection{NRQCD Calculation}
\label{subsec:chromolat}

The leading spin-dependent term in the effective action for NRQCD is
\begin{equation}
 \Gamma^{\NRQCD}= c_4Z_\sigma^{\NR} g\psi^\dagger\frac{\boldsigma\cdot\boldB}{2M}\psi+\ldots\ ,
\end{equation}
which, after renormalization, becomes
\begin{equation}
\Gamma^{\NRQCD} = 
  c_4Z_\sigma^{\NR}Z_2^{\NR}Z_m^{\NR}\psi_R^\dag\frac{\vec{\sigma}\cdot\vec{B}}{2M_R}\psi_R\ .
\end{equation}  
Since the chromomagnetic operator is present in the NRQCD action at tree level,
$Z_\sigma$ is of the form $Z_\sigma^{\NR}=1+Z_\sigma^{\NR,(1)}\alpha_s$. 
Requiring that the anomalous chromomagnetic moment to be equal 
in QCD and NRQCD, we obtain the matching condition
\begin{equation}
c_4 Z_\sigma^{\NR} Z_2^{\NR} Z_m^{\NR} = 1 + Z_\sigma,
\end{equation} 
which at tree level and one-loop order yields
\begin{eqnarray}
c_4^{(0)} &=& 1\;, \\ \nonumber
c_4^{(1)} &=& b_\sigma^{(1)}-Z_\sigma^{\NR,(1)}-Z_2^{\NR,(1)}-Z_m^{\NR,(1)}\;.
\end{eqnarray} 
Note that mass renormalization has to be included in NRQCD since the
chromomagnetic operator is now included at tree-level.

All of the renormalization constants involved are UV finite and consist of
two distinct contributions:
besides the ordinary diagrammatic contributions (which we denote simply as $Z$),
there are also contributions from mean-field improvement ($Z^{tad}$).
The NRQCD diagrams contributing to $Z_\sigma^{\NR,(1)}$ are shown in Figure 
\ref{fig:vertexdiags} (a)-(e). Note that diagrams (c)-(e) receive contributions
not only from lattice artifacts, but also from higher-adicity vertices
that are also present in continuum NRQCD.

Having once coded the Feynman diagrams in Figure \ref{fig:vertexdiags} using
the \textsc{HPsrc} package, we can repeat the numerical evaluation
of these diagrams for each NRQCD action of interest by using the \textsc{HiPPy}
package to automatically generate Feynman rules for that action along
with the Symanzik-improved gluonic action.

By using the residual gauge-invariance of the effective potential in BFG,
the contribution to the negative of the effective action from the diagrams
can be restricted to the form
{\bacol
\begin{eqnarray}
-\delta\Gamma_{\psib\psi A_i}(p,q)=&& \nonumber\\
 -Z_\sigma^{\NR,(1)} g\psi(p+q)^\dagger && \frac{i\epsilon_{ijk}\sigma_jq_kA_i(q)}{2M}
\psi(p)+\ldots\ ,
\end{eqnarray}
}
where $i=1,2,3$ and terms of higher order in the external momentum $q$ have been
omitted. Without BFG, non-gauge invariant operators would be generated as
counter-terms, greatly increasing the difficulty of the calculation
and the subsequent use of the improved action in simulations.
Using the automatic differentiation
\cite{Hippel06,Hippel10}
and spinor manipulation facilities built into the \textsc{HPsrc} package,
the coefficient of interest can be isolated as
\begin{equation}
Mi\frac{\partial}{\partial p_1}Tr\left(\delta\Gamma_{\psib\psi A_2}(p,0)\sigma_3\right)= 
g Z_\sigma^{\NR,(1)}\ ,
\end{equation}
where $\mathbf{p}=0$, with $p_0$ taken such that the quark is on-shell
using the lattice equation of motion as implemented in the \textsc{HPsrc}
package.

We integrate over the temporal component $k_4$ of the loop momentum $k$ by
contour integration over the unit circle $w=e^{ik_4}$ in the complex plane.
The positions of the poles in the integrand arising from the gluon and quark
propagators in the Feynman diagram are functions of the spatial
loop momentum $\bk$. To ensure that no poles cross the contour of integration
as $\bk$ varies, we change the contour to be a circle of radius $r$,
choosing $r$ such that the contour lies exactly half-way between the
outermost interior pole and the innermost exterior pole.
Since there are no physical intermediate states and thus no 
branch cut in any of the diagrams this strategy is always possible
\cite{LewThesis}.
Since it is known that the poles of full NRQCD and improved gluonic
actions always lie further away from the contour than their unimproved
counterparts
\cite{Horgan:2009},
it is therefore safe to reduce the effort by locating the poles
corresponding to the unimproved action and shift the contour
appropriately.

\begin{figure}[h]
\centering
\includegraphics[width=0.70\linewidth,clip=true]{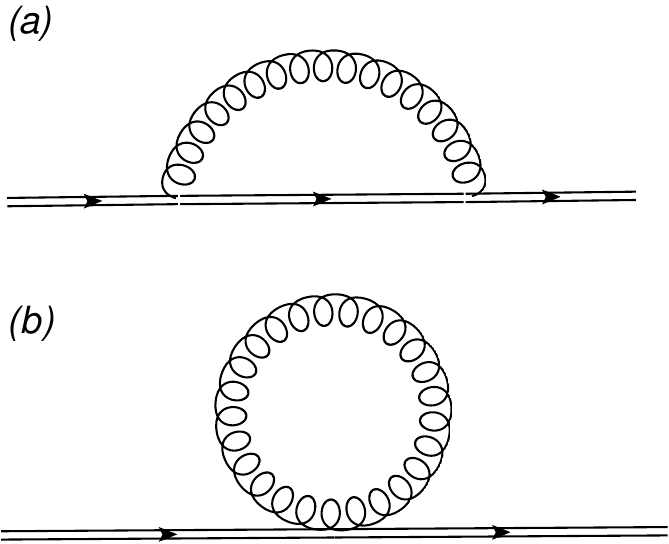}
\caption{Quark two-point function Feynman diagrams.
         Referred to as (a) rainbow and (b) tadpole diagrams.}
\label{fig:quark2pointdiag}
\end{figure}

The renormalization constants
\begin{eqnarray}
Z_2^{\NR,(1)}&=&1+\textrm{Re}{\Sigma(0)}+\textrm{Im}{\frac{\partial\Sigma}{\partial p_4}}\bigg|_{p=0}\;,\\
Z_m^{\NR,(1)}&=&1+M\textrm{Re}{\frac{\partial^2\Sigma}{\partial p_3^2}}\bigg|_{p=0}-
                \textrm{Im}{\frac{\partial\Sigma}{ \partial p_4}}\;,
\end{eqnarray}
are determined from the quark self-energy
$-\Gamma_{\psib\psi}(-p,p)=\Sigma(p)$,
which is given by the diagrams shown in Figure \ref{fig:quark2pointdiag}.

Both $Z_2^{\NR,(1)}$ and $Z_\sigma^{\NR,(1)}$ contain logarithmic IR divergences.
For our purposes, it sufficient to evaluate their sum for which the IR logarithm 
is known from analytical results. We can therefore determine
the IR finite diagrammatic contribution $\widetilde Z_X$ from fitting
our numerical results with the form
\begin{equation}
Z_\sigma^{\NR,(1)}+Z_2^{\NR,(1)}=\widetilde Z_\sigma^{\NR,(1)}+ \widetilde Z_2^{\NR,(1)}+\frac{3}{2\pi}\log{\mu a}.
\end{equation}
We evaluate the diagrams for $10^{-8}<\mu<10^{-6}$ such that any $(\mu a)^2$
lattice artifacts can be neglected in the fit. The logarithmic IR divergence
combines with the IR logarithm from the QCD result above to yield an overall
logarithmic contribution $-\frac{3\alpha_s}{2\pi} \log(M a)$ to $c_4^{(1)}$.  

$Z_\sigma^{\NR,(1)}$ and $Z_m^{\NR,(1)}$ also receive corrections from
tadpole improvement, whereas there are no tadpole corrections to $Z_2^{\NR,(1)}$.
Using a \textsc{Mathematica}
\cite{mathematica}
notebook developed in the course of earlier related work
\cite{EikeThesis,LewThesis},
the tadpole corrections are determined by symbolically substituting
mean field corrections, $U\rightarrow U/u_0$ with $u_0=1-\alpha_s u_0^{(2)}$,
into the NRQCD action. The tadpole corrections are dependent
on the details of the NRQCD action used, in particular also on the value
of the stability parameter $n$.

For the full $v^4$ NRQCD action we find
\begin{equation}
Z_m^{\rm tad,(1)} = -\left(\frac{2}{3} +\frac{3}{(Ma)^2}\right)u_0^{(2)}.
\end{equation}
The tadpole contribution to $Z_\sigma^{\NR,(1)}$ arises from two sources:
the application of mean-field improvement to the improved field-strength tensor,
and the cross-multiplication of the tree-level $\vec{\sigma}\cdot\vec{B}$
term with the tadpole corrections terms in $H_0$
\cite{PhysRevD.72.094507}.
The overall result for the full $v^4$ NRQCD action is
\begin{equation}
Z_\sigma^{\rm tad, (1)} = \left(\frac{13}{3}+\frac{13}{4Ma}-\frac{3}{8 n (Ma)^2}-\frac{3}{4(Ma)^3}
\right)\;
u_0^{(2)}\;.
\end{equation}
We take the one-loop contribution to the Landau mean link to be
$u_0^{(2)} = 0.750$
\cite{Nobes:2001tf}.
%

%%%%%%%%%%%%%%%%%%%%%%%%%%%%%%%%%%%%%%%%%%%%%%%%%%%%%%%%%%%%%%%%%%%%%%%%%%%%%%
\subsection{Results}
\label{subsec:chromores}

The final result for the radiative correction to $c_4$ is given by
\begin{eqnarray}
c_4^{(1)} &=&\textstyle \frac{13}{6\pi}-\widetilde Z_\sigma^{\NR,(1)}-\widetilde Z_2^{\NR,(1)}-
\widetilde Z_m^{\NR,(1)}\nn \\ 
&-&\textstyle Z_m^{\rm tad,(1)} - Z_\sigma^{\rm tad, (1)} -\frac{3}{2\pi} 
\log Ma. 
\label{eq:c4}
\end{eqnarray}
Results for several different NRQCD actions,
all used with the tree-level Symanzik gluon action,
are summarized in Tables \ref{tab:sigmaBresults1}-\ref{tab:sigmaBresults3}. 
The values of $Ma$ were chosen to correspond to the lattice
spacings for lattices used by the HPQCD collaboration
\cite{PhysRevD.72.094507}.

We note that the tadpole corrections to $Z_\sigma$ and $Z_m$
in each case very nearly cancel the diagrammatic contribution,
demonstrating that they are working as intended by reducing
the coefficient of $\alpha_s$ in the perturbative series.
Without such a tadpole correction we find $c_4^{(1)}\approx 4$, meaning that
the viability of the perturbative expansion would be highly questionable.

On the other hand, the radiative corrections for the full $v^4$ NRQCD action
with stability parameter for $n=2$ or $n=4$ differ very little;
in the case of the full spin $v^6$ NRQCD action it appears that for small values
of $Ma$ the correction increases slightly, but for larger values of $Ma$ 
the corrections are very similar to the $v^4$ case.
These results suggest that the radiative corrections for the
chromomagnetic operator are relatively independent of the details
of the NRQCD action.

Note that for all actions and ranges of $Ma$ the total correction is positive:
the constant part of the correction is larger than the (negative) logarithmic
contribution. This refutes the claims of Penin
\cite{Penin09},
who does not include a calculation of the $Ma$-independent
constant contribution to $c_4^{(1)}$ which is responsible for
$c_4^{(1)}$ being positive.

\begin{figure}[h]
\centering
\includegraphics[width=0.90\linewidth,clip=true]{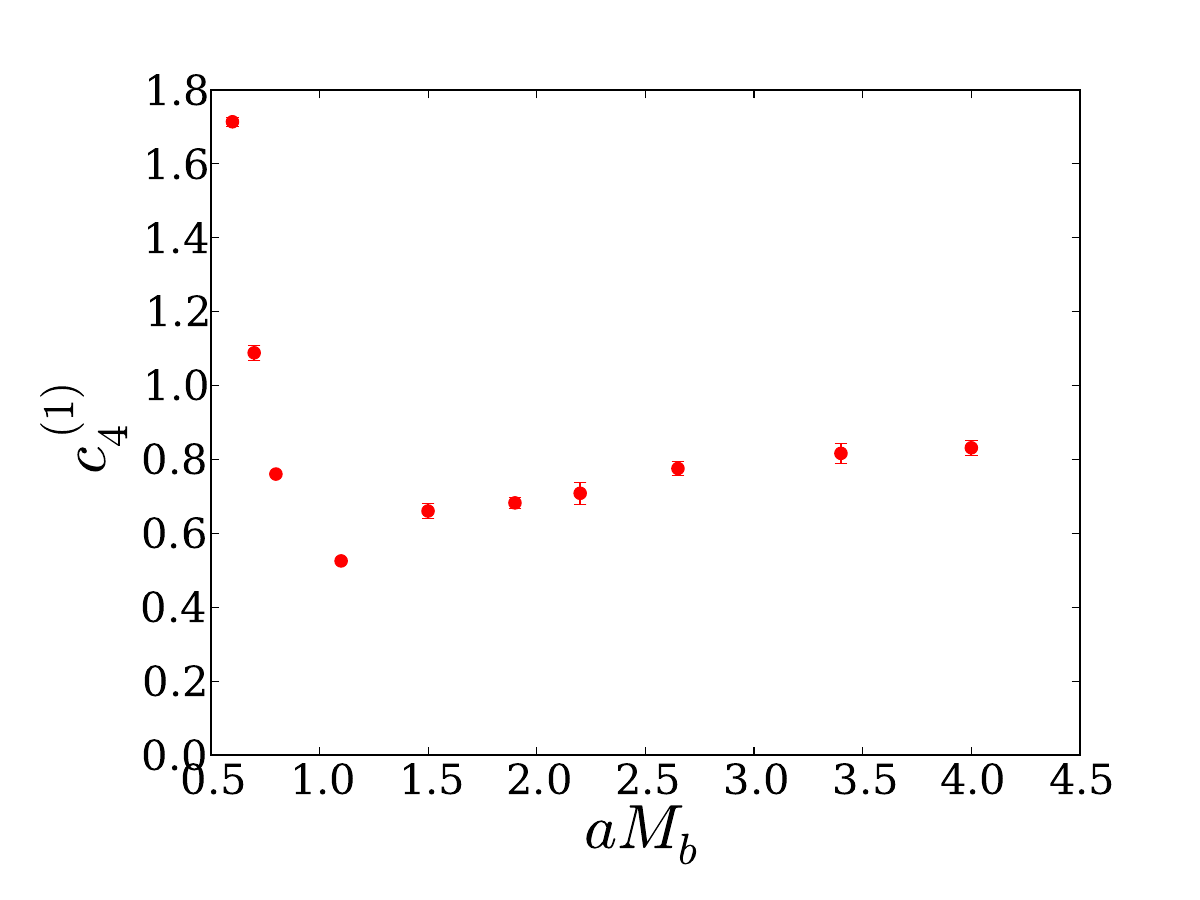}
\caption{Lattice spacing dependence for the chromomagnetic operator correction,
         full $O(v^4)$ NRQCD $n=4$ action.}
\label{fig:n4Mdep}
\end{figure}

In Figure \ref{fig:n4Mdep} we plot the dependence of the radiative
correction $c_4^{(1)}$ against $Ma$ for full NRQCD $n=4$.
While the expected divergence in the $Ma\to0$ limit can be seen,
for values of $2<Ma<4$ the correction varies only slowly,
demonstrating that for the range of lattice spacings used by the HPQCD
collaboration the perturbative improvement corrections are under control.

\begin{table}[h]
\centering
\begin{tabular}{|c|ccccc|}
\hline
$Ma$ & 1.95 &~~& 2.8 &~~& 4.0 \\\hline\hline
$\widetilde Z_\sigma^{(1)}+\widetilde Z_2^{(1)}$ & -5.164(7) && -4.913(6) && -4.739(6) \\
$\widetilde Z_m^{(1)}$ & 1.512(1)  && 1.022(3) && 0.723(2) \\
$Z_\sigma^{\rm tad, (1)}$ & 4.387 && 4.077 && 3.841 \\
$Z_m^{\rm tad, (1)}$ & -1.092 && -0.787 && -0.641  \\
$c_4^{(1)}$ & 0.728(7) && 0.799(7) && 0.842(6) \\\hline
\end{tabular}
\caption{Renormalization parameters for the chromomagnetic term, full $v^4$ NRQCD $n=2$}
\label{tab:sigmaBresults1}
\end{table}

\begin{table}[h]
\centering
\begin{tabular}{|c|ccccc|}
\hline
$Ma$ & 1.9 &~~& 2.65 &~~& 3.4 \\\hline\hline
$\widetilde Z_\sigma^{(1)}+\widetilde Z_2^{(1)}$ & -5.17(1) && -4.94(1) && -4.80(2) \\
$\widetilde Z_m^{(1)}$ & 1.56(1)  && 1.08(1) && 0.84(1) \\
$Z_\sigma^{\rm tad, (1)}$ & 4.43 && 4.129 && 3.95 \\
$Z_m^{\rm tad, (1)}$ & -1.12 && -0.82 && -0.69  \\
$c_4^{(1)}$ & 0.68(1) && 0.78(2) && 0.82(3) \\\hline
\end{tabular}
\caption{Renormalization parameters for the chromomagnetic term, full $v^4$ NRQCD $n=4$}
\label{tab:sigmaBresults2}
\end{table}

\begin{table}[h]
\centering
\begin{tabular}{|c|ccccc|}
\hline
$Ma$ & 1.9 &~~& 2.65 &~~& 3.4 \\\hline\hline
$\widetilde Z_\sigma^{(1)}+\widetilde Z_2^{(1)}+\widetilde Z_m^{(1)}$ & -4.33(8) && -4.16(7) && -4.16(5) 
\\
$Z_\sigma^{\rm tad, (1)}+Z_m^{\rm tad, (1)}$ & 3.93 && 3.63 && 3.45 \\
$c_4^{(1)}$ & 0.78(8) && 0.76(7) && 0.83(5) \\\hline
\end{tabular}
\caption{Renormalization parameters for the chromomagnetic term, full spin $v^6$ NRQCD $n=4$}
\label{tab:sigmaBresults3}
\end{table}

%%%%%%%%%%%%%%%%%%%%%%%%%%%%%%%%%%%%%%%%%%%%%%%%%%%%%%%%%%%%%%%%%%%%%%%%%%%%%%%%
%%%%%%%%%%%%%%%%%%%%%%%%%%%%%%%%%%%%%%%%%%%%%%%%%%%%%%%%%%%%%%%%%%%%%%%%%%%%%%%%
\section{The Darwin Operator}
%%%%%%%%%%%%%%%%%%%%%%%%%%%%%%%%%%%%%%%%%%%%%%%%%%%%%%%%%%%%%%%%%%%%%%%%%%%%%%%%
%%%%%%%%%%%%%%%%%%%%%%%%%%%%%%%%%%%%%%%%%%%%%%%%%%%%%%%%%%%%%%%%%%%%%%%%%%%%%%%%
\label{sec:darwin}

The vertex correction diagrams of Figure \ref{fig:vertexdiags} also renormalize
the $\boldD\cdot\boldE$ operator (conventionally called the Darwin term in analogy
with atomic physics) in the effective action. The primary effect of this
operator is to change the effective potential for the wavefunction at the
origin; since only states with $L=0$ have a non-vanishing wavefunction at the
origin, this results in an energy shift for S-wave states.

In the same manner as for the chromomagnetic operator, we can tune the
coefficient $c_2$ of the Darwin term so as to correct for the difference
between the loop corrections of QCD and NRQCD. Previously, these corrections
were unknown, and the coefficient $c_2$ has been set equal to one in most
non-perturbative simulations so far. While non-perturbative studies
have shown a comparatively mild dependence of energy levels on varying the
value of $c_2$
\cite{Dowdall:2011wh},
a precise determination of the radiative corrections to $c_2$ is a logical
continuation of our improvement programme.

%%%%%%%%%%%%%%%%%%%%%%%%%%%%%%%%%%%%%%%%%%%%%%%%%%%%%%%%%%%%%%%%%%%%%%%%%%%%%%
\subsection{Continuum QCD Calculation}
\label{subsec:darwincont}

In continuum QCD, we need to consider the $q^2$-dependence of the effective
action,
\begin{equation}
 \Gamma^{QCD}= \Psib F_1(q^2)\gA\Psi+\Psib\frac{F_2(q^2)\sigma^{\mu\nu}F_{\mu\nu}}{2m}\Psi+\ldots\,,
\end{equation}
where for the chromomagnetic case we had
$F_1(0)=Z_{1F}(=Z_2^{-1})$ and $F_2(0)=Z_\sigma$.
Upon performing the non-relativistic reduction and isolating the terms
containing the time component $A_0$ of the gauge field, we find
\begin{eqnarray}
 \Gamma^{QCD}=&&F_1(q^2)\psi^\dagger\left[gA_0-\frac{g}{8m^2}q^2A_0+\ldots\right]\psi \nonumber\\
+&&F_2(q^2)\psi^\dagger \left[-\frac{g}{4m^2}q^2A_0+\ldots\right]\psi,
\end{eqnarray}
which, after renormalization, gives the Darwin term as
{\bacol
\begin{eqnarray}
&& \Gamma^{QCD}= \nonumber\\
&&~~\left[1-8m^2 F^\prime_1(0)+2F_2(0)\right]\psi^\dagger_R\left[-\frac{g}{8m_R^2}q^2A_0\right]
\psi_R+\ldots\ . \nonumber\\
\end{eqnarray}
}
Note that $F_2$ contributes through the non-relativistic reduction,
whereas $F_1$ only contributes through its expansion around $q^2=0$,
since the wavefunction renormalization cancels any contribution from $F_1(0)$.
Again, we do not have to include mass renormalization since the renormalized
mass $m$ naturally appears when using the non-relativistic reduction between
on-shell spinors.

$F_2(0)$ has already been computed for the chromomagnetic case.
For the contribution from the abelian diagram of Figure \ref{fig:vertexdiags}(a),
one finds
\begin{equation}
F^\prime_{1a}(0)=\frac{\alpha_s}{8 m^2}\left[-\frac{1}{6\pi}-\frac{4}{9\pi}\log{\mu/m}\right],
\end{equation}
and for the non-abelian diagram of Figure \ref{fig:vertexdiags}(b) we may take
the derivative of the analytical calculation carried out for the chromomagnetic
term to obtain
{\bacol
\begin{eqnarray}
F^\prime_{1b}(0)=&& \nonumber\\
\frac{\alpha_s}{8 m^2} && \left[\frac{m^2}{\mu^2\pi}+\frac{7m}{4\mu}+\frac{11}{2\pi}+
\frac{9}{\pi}\log{\mu/m}\right].
\end{eqnarray}
}
The total contribution to the continuum QCD Darwin term is then
{\bacol
 \begin{eqnarray}
Z_D=&& \label{eq:ZDcont}\\
1+\alpha_s&&\left[-\frac{m^2}{\mu^2\pi}  -\frac{7m}{4\mu}-\frac{1}{\pi}+(-\frac{6}{\pi}+\frac{4}{9
\pi})\log{\mu/m}\right]\;,\nonumber
\end{eqnarray}
}
where as before, we use a gluon mass $\mu$ as the IR regulator.
We note, however, that the case of the Darwin term is more subtle since
there are power-law IR divergences which will have to match and cancel
with corresponding IR divergences on the NRQCD side.

%%%%%%%%%%%%%%%%%%%%%%%%%%%%%%%%%%%%%%%%%%%%%%%%%%%%%%%%%%%%%%%%%%%%%%%%%%%%%%
\subsection{NRQCD Calculation}
\label{subsec:darwinlat}

The Darwin term in the NRQCD effective action is given by
\begin{equation}
 \Gamma^{\NRQCD}= -c_2Z_D^{\NR} g\psi^\dagger\frac{q^2A_0}{8M^2}\psi+\ldots
\end{equation}
where again $Z_D^\NR$ has a tree-level contribution,
$Z_D^{\NR}=1+Z_D^{\NR,(1)}\alpha_s+\ldots$. After renormalization we obtain,
\begin{equation}
  \Gamma^{\NRQCD}= -c_2Z_D^{\NR}(Z_m^{\NR})^{2}Z_2^{\NR}g\psi_R^\dagger\frac{q^2A_0}{8M_R^2}\psi_R+\ldots\;. 
\end{equation}
Requiring the coefficients of the Darwin term to be equal in QCD and NRQCD,
we find the matching condition
\begin{equation}
c_2 Z_D^{\NR} Z_2^{\NR} (Z_m^{\NR})^2 = Z_D,
\end{equation}
which at tree level and one-loop order gives
\begin{eqnarray}
c_2^{(0)} &=& 1\;, \nonumber \\
c_2^{(1)} &=& Z_D^{(1)}-Z_D^{\NR,(1)}-Z_2^{\NR,(1)}-2Z_m^{\NR,(1)}\;.
\end{eqnarray}
As in the chromomagnetic case, the diagrammatic contributions, labelled as $Z$,
must be supplemented by the corresponding mean-field corrections, $Z^{tad}$.

Again, the symmetries of the effective action in BFG restrict the form of
the diagrammatic contributions to
\begin{equation}
-\delta\Gamma_{\psib\psi A_0}(p,q)= -Z_D^{\NR,(1)} g\psi(p+q)^\dagger\frac{q^2A_0(q)}{8M}\psi(p)+\ldots \,.
\end{equation}
Working in the Breit frame ($p_i=-q_i/2$) for ease of implementation,
we isolate the renormalization constant 
{\bacol
\begin{eqnarray}
Z_D^{\NR,(1)}=&& \nonumber\\
-\frac{4M^2}{3}\sum_i&&\frac{\partial^2}{\partial q_i^2}\delta\Gamma_{\psib\psi 
A_0}(0,0)-M\frac{\partial}{\partial p_0}\delta\Gamma_{\psib\psi A_0}(0,0),\nonumber\\
\end{eqnarray}
}
where the second term arises from the on-shell condition of the incoming and
outgoing quarks. We note that the final result is of course independent of the
choice of frame used.

By taking advantage of the modular structure of the \textsc{HiPPy} and
\textsc{HPsrc} packages, we are able to reuse the same code that we used
for matching the chromomagnetic operator by merely changing the incoming
gauge field Lorentz index to isolate the $A_0$ component, and taking the
trace of the diagram to isolate the implicit Dirac unit matrix in front
of the Darwin operator. The pole structure for each individual diagram
remains as in the previous calculations.

The presence of severe IR divergences in some of the diagrams makes it
necessary to analytically identify and subtract those divergences,
leaving only the IR finite piece $\widetilde Z$ to be computed numerically.
For the sake of brevity, the superscript $NR$ is suppressed in the remainder
of this section except where it is necessary to avoid confusion.

For the diagram in Figure \ref{fig:vertexdiags}(a) we obtain
\begin{equation}
Z_D^{a,(1)}=\widetilde Z_D^{a,(1)}+\left[\frac{4}{9\pi} - 
                              \left(\frac{1}{6\pi}\right)\right]\log{\mu a},
\label{eq:ZDa}
\end{equation}
where the IR divergence in round brackets arises from the inclusion of
the tree-level Darwin term in the action. We therefore would expect this
divergence to be cancelled by an IR divergence in the wavefunction
renormalization times the tree-level operator. Since the IR divergences
are calculated analytically we can use a constrained fit to find the
constant $\mu a$-independent contribution, $\widetilde Z_D^{(a),(1)}$. 

\begin{figure}[t]
\centering
\includegraphics[width=0.90\linewidth,clip=true]{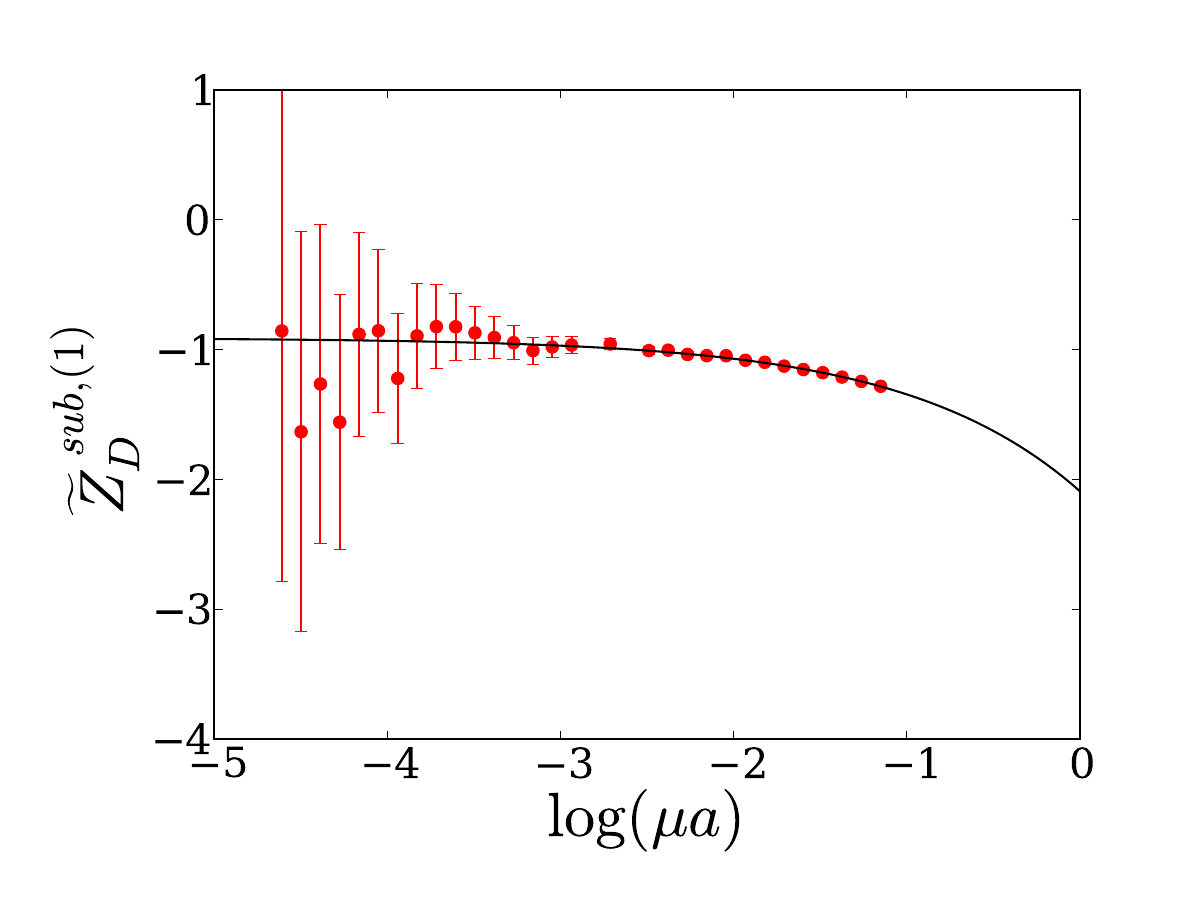}
\caption{Sample fit for $\widetilde Z_D^{sub, (1)}$. $Ma=1.9$.
         We obtain $-0.91(1)-1.18(5)\mu a$ with $\chi^2/d.o.f.=0.21$.}
\label{fig:darwinsubplot}
\end{figure}

For the diagram in Figure \ref{fig:vertexdiags}(b) we have
\begin{equation}
Z_D^{(b),(1)}=\widetilde Z_D^{(b),(1)}+D(Ma)\log{\mu a}-\frac{7M}{4\mu}-\frac{M^2}{\pi\mu^2},
\end{equation}
where $D(Ma)$ is an action-specific coefficient containing lattice artifacts.
The leading continuum power-law divergences are removed by the subtraction
function
\begin{widetext}
\begin{eqnarray}
I^{sub}(\mu)&=&(4\pi)(8M^2)\int \frac{d^4k}{(2\pi)^4}\left[ \frac{\frac{7}{8M}}{(k^2+\mu^2)^2(ik_0+k^2/2M)}+
\frac{(3\mu^2)}{(k^2+\mu^2)^4}\right]\nn\\
&=&\widetilde Z^{sub, (1)}+\frac{7M}{4\mu}+\frac{21}{4\pi}\log{\mu a}+\frac{M^2}{\pi\mu^2} + P^{sub}(\mu a)\; ,
\label{eqn:Isub}
\end{eqnarray}
\end{widetext}
where $P^{sub}(\mu a)$ is a polynomial in $\mu a$ with $P^{sub}(0) = 0$.
This subtraction function is chosen to cancel the leading $M^2/\mu^2$
and $M/\mu$ IR divergences pointwise and also contains a continuum-like
IR log divergence which is shown in Eq. (\ref{eqn:Isub}).
We calculate $\widetilde Z^{sub, (1)}$ from a fit to $I^{sub}(\mu)$,
using values of the cutoff in the range $10^{-3}<\mu<10^{-1}$ and fitting
the results to a polynomial in $\mu a$. In Figure \ref{fig:darwinsubplot}
we show a typical fit for the subtraction function giving the required
renormalization constant $\widetilde Z^{sub}$. From the figure we note
that the magnitude of the error increases as $\mu$ decreases and the cost
of computation correspondingly increases. 

In addition to containing continuum IR divergences, the graph in
Figure \ref{fig:vertexdiags}(b) contains additional artifact logarithmic
IR divergences in NRQCD which must ultimately cancel against similar
contributions from the graphs in Figure \ref{fig:vertexdiags}(c-f).
We calculate these diagrams, namely the swordfish, algae, and ankh diagrams,
and add their contribution to that of the subtracted diagram in
Figure \ref{fig:vertexdiags}(b) to compute the overall logarithmic
IR divergence which should not (and indeed does not) contain a lattice
artefact IR logarithm.   

For the swordfish, algae and ankh diagrams we write
\begin{equation}
Z_D^{c-f, (1)}=\widetilde Z_D^{c-f, (1)}+D^\prime(Ma)\log{\mu a},
\end{equation}
where $D^\prime(Ma)$ is the action-specific coefficient of the lattice
artefact logarithmic IR divergence. We find that
\begin{equation}
D(Ma)+D^\prime(Ma)=-\frac{6}{\pi} + \left(\frac{3}{2\pi}\right),
\label{eq:ZDb}
\end{equation}
which is independent of lattice artifacts, as expected. As before, the term
in round brackets arises from the inclusion of the tree-level Darwin term
in the NRQCD action. The graphs in Figure \ref{fig:vertexdiags}(b-f)
are combined with the subtraction function in Eq. (\ref{eqn:Isub}) and the
integral computed numerically. Combining the diagrams in Figure
\ref{fig:vertexdiags}(b-f) we compute, as a function of $\mu a$,
the combination
\begin{eqnarray}
Z_D^{b,(1)}+I^{sub}-\widetilde Z^{sub, (1)}+Z_D^{c-f, (1)}=&& \nonumber\\
\widetilde Z^{b, (1)}+\widetilde Z^{c-f, (1)}+\frac{3}{4\pi}\log{\mu a}+P^{b-f}(\mu a)\; ,~~~&&
\end{eqnarray}
where $P^{b-f}(\mu a)$ is a polynomial in $\mu a$ with $P^{b-f}(0)=0$, and
the fit to the parameterization with the coefficient of the logarithmic
IR divergence constrained to its analytic value gives a much better 
estimate for the required renormalization constants.  

The presence of the tree-level Darwin term in the NRQCD action means we
have also to include the contribution from the wavefunction renormalization  
\begin{equation}
Z_2^{NR, (1)}=\widetilde Z_2^{(1)}-\frac{4}{3\pi}\log{\mu a}.
\end{equation}
As expected, the IR logarithm in $Z_2$ does indeed cancel the sum of
the logarithmic IR divergences displayed in round brackets in
Eqs. (\ref{eq:ZDa}) and (\ref{eq:ZDb}).

The mass renormalization is finite and so can be simply added to the result.

As for the chromomagnetic term, we must also include tadpole contributions,
which in this case come from three sources: the improvement of the field
strength tensor, the improvement of $\nabla^{(\pm)}$, and from
cross-multiplication with tadpole-improved links in the other parts of the
NRQCD action. Altogether, this gives
\begin{equation}
\delta Z_D^{tad}=\left(\frac{16}{3}-\frac{3}{4M^3}-\frac{3}{32M^2}+\frac{13}{4M}\right)u_0^{(2)}\; ,
\end{equation}
which is the same as the tadpole correction to the chromomagnetic term
plus $1$ from the correction of $\nabla^{(\pm)}$. The mass renormalization
tadpole is the same as used in the chromomagnetic calculation. 

All lattice artefact IR divergences cancel internally within the
NRQCD calculation and the continuum IR log divergences match between the
continuum and lattice NRQCD calculations. Using the result for the continuum
contribution in Eq. (\ref{eq:ZDcont}) and summing all lattice NRQCD
contributions, we find the one-loop improvement coefficient for the NRQCD
Darwin term to be
\begin{widetext}
 \begin{equation}
c_2^{(1)}=-\frac{1}{\pi}-\widetilde Z^{b, (1)}-\widetilde Z^{c-f, (1)}-\widetilde Z^{a, (1)}- 
Z_D^{tad, (1)}-\widetilde Z_2^{(1)}-2Z_m^{(1)}-2Z_m^{tad, (1)}+\frac{50}{9\pi}\log{Ma}\; ,
\end{equation}
\end{widetext}
where again the logarithmic divergences in NRQCD and QCD have combined to
give a term proportional to $\log{Ma}$.

%%%%%%%%%%%%%%%%%%%%%%%%%%%%%%%%%%%%%%%%%%%%%%%%%%%%%%%%%%%%%%%%%%%%%%%%%%%%%%
\subsection{Results}
\label{subsec:darwinres}

In Tables \ref{Darwinn4} and \ref{Darwinn4v6} we give results for the
full $v^4$ NRQCD action and the full spin $v^6$ NRQCD action respectively,
both with stability parameter $n=4$. For brevity, we use the notation
\begin{equation}
\widetilde Z^{a-f, (1)} = \widetilde Z^{a, (1)} + \widetilde Z^{b, (1)} + \widetilde Z^{c-f, (1)}.
\end{equation}
Again, we find that the tadpole correction cancels the major part of the
diagrammatic contribution to $c_2^{(1)}$, indicating improved convergence
of the resulting perturbative series. This effect is expected and reinforces
the efficacy of including tadpole improvement.

Numerical simulations
\cite{Dowdall:2011wh}
show that using a value of $c_2=1.25$ shifts the energy of the $\Upsilon(1S)$
state by less than one percent compared with the tree-level value $c_2=1$,
indicating that the radiative corrections are well under control.

\begin{table}[h]
\centering
\begin{tabular}{|c| c c c c c|}
\hline
$Ma$ & $\widetilde Z^{a-f, (1)}$ & $ Z_D^{tad, (1)}$ & $c_2^{(1)}$ & $\alpha_V$ & $c_2$ \\ [0.5ex]\hline\hline 
1.9 &-5.95(8)&5.18 & 1.33(8) & 0.22 & 1.29(2)  \\
2.65 &-3.71(10) &4.88 & 0.08(10) & 0.25 & 1.02(3)\\
3.4 &-1.73(12)&4.69 & -1.20(12)& 0.27 & 0.68(4)\\ [1ex]
\hline
\end{tabular}
\caption{Renormalization parameters for the Darwin term, full $v^4$ NRQCD $n=4$}
\label{Darwinn4}
\end{table}
 \begin{table}[h]
\centering
\begin{tabular}{|c| c c c c c|}
\hline
$Ma$ & $\widetilde Z^{a-f, (1)}$ & $ Z_D^{tad, (1)}$ & $c_2^{(1)}$ & $\alpha_V$ & $c_2$ \\ [0.5ex]\hline\hline 
1.9 &-5.83(8)&5.18 & 1.32(9) & 0.22 & 1.29(2)  \\
2.65 &-3.64(10) &4.88 & 0.10(10) & 0.25 & 1.03(3)\\
3.4 &-1.67(11)&4.69 & -1.23(11)& 0.27 & 0.67(4)\\ [1ex]
\hline
\end{tabular}
\caption{Renormalization parameters for the Darwin term, full spin $v^6$ NRQCD $n=4$}
\label{Darwinn4v6}
\end{table}

%%%%%%%%%%%%%%%%%%%%%%%%%%%%%%%%%%%%%%%%%%%%%%%%%%%%%%%%%%%%%%%%%%%%%%%%%%%%%%%%
%%%%%%%%%%%%%%%%%%%%%%%%%%%%%%%%%%%%%%%%%%%%%%%%%%%%%%%%%%%%%%%%%%%%%%%%%%%%%%%%
\section{Four-fermion interactions}
%%%%%%%%%%%%%%%%%%%%%%%%%%%%%%%%%%%%%%%%%%%%%%%%%%%%%%%%%%%%%%%%%%%%%%%%%%%%%%%%
%%%%%%%%%%%%%%%%%%%%%%%%%%%%%%%%%%%%%%%%%%%%%%%%%%%%%%%%%%%%%%%%%%%%%%%%%%%%%%%%
\label{sec:4f}

At the one-loop level, calculations in NRQCD must also take account of
four-fermion interactions in the NRQCD action that are needed to match
$\overline{Q}Q\to \overline{Q}Q$ scattering processes between QCD and NRQCD.
While so far we have emphasized the r\^ole of the effective action
in BFG, the matching condition must ultimately equate physical, on-shell
matrix element, which in general will include contributions from 1PR
diagrams, and so it is not necessarily possible to match just the effective
actions term by term.

%%%%%%%%%%%%%%%%%%%%%%%%%%%%%%%%%%%%%%%%%%%%%%%%%%%%%%%%%%%%%%%%%%%%%%%%%%%%%%
\subsection{Formalism}
\label{subsec:4fform}

The radiatively generated four-fermion interactions that need to be added
to the NRQCD action can be written as four-fermion contact operators in a
(covariant) derivative expansion. Here we consider only the lowest-order
terms in this expansion, namely those without derivatives.
One obvious set of operators which bear a close relationship to the
$\overline{Q}Q\to \overline{Q}Q$ diagrams calculated, is given by
{\bacol
\begin{eqnarray}
 S^{NRQCD} =&& \nonumber\\ 
a_{8}\frac{\alpha_s g^2}{M^2}(\chi^\dagger && T_a^T\chi)(\psi^\dagger T_a\psi) 
+ a_{1} \frac{\alpha_s g^2}{M^2}(\chi^\dagger\chi)(\psi^\dagger\psi)\nonumber\\
+b_{8}\frac{ \alpha_s g^2}{M^2}(\chi^\dagger && \sigma^*T_a^T\chi)(\psi^\dagger\sigma T_a\psi) 
+ b_{1} \frac{\alpha_s g^2}{m^2}(\chi^\dagger\sigma^*\chi)(\psib\sigma \psi)\; . \nonumber\\
\label{eq:Sab}
\end{eqnarray}
}
We will refer to the operators with coefficients $a_{1}$ and $b_{1}$
as singlet-exchange operators.
The operators with coefficients $a_{8}$ and $b_{8}$, which we will refer to
as octet-exchange operators, give corrections to processes involving
single gluon exchange at tree level. This relation will be useful later when
discussing corrections to the QCD Coulomb force.
The spin-independent and spin-dependent operators have coefficients $a_i$
and $b_i,~i=1,8$, respectively.
Only $b_1$ and $b_8$ will contribute to the hyperfine splitting, as can be
seen by choosing the more conventional set of operators given by
\cite{Labelle:1997,PhysRevD.46.4052}
\begin{eqnarray}
 S^{\NRQCD} {}={} d_1\frac{\alpha_s^2 }{M^2}(\psi^\dagger\chi^*)(\chi^T\psi) +d_2\frac{\alpha_s^2}{M^2}(\psi^
\dagger\sigma\chi^*)(\chi^T\sigma\psi)\nn\\
 +d_3\frac{\alpha_s^2}{M^2}(\psi^\dagger T_a\chi^*)(\chi^T T_a\psi) +d_4\frac{\alpha_s^2}{M^2}(\psi^\dagger
\sigma T_a\chi^*)(\chi^T\sigma T_a\psi)\; .\nonumber\\
\label{eq:Sd}
\end{eqnarray}
Here, we choose the $\chi$ field for the antiquark to transform according to the conjugate $\overline{1/2}$
representation of spin and the $\overline{3}$ representation of colour generated, respectively, by $-\sigma^*$ and
$-T^*$. Such s-channel operators make explicit their effects on different meson states; this will be useful in 
later discussions of the hyperfine splitting.

These two sets of operators in Eqs. (\ref{eq:Sab}), (\ref{eq:Sd}) are related by Fierz transformations.
Considering the action of the operators on
colour octet ($\frac{1}{\sqrt{3}}T_b$) and
singlet states ($\frac{1}{\sqrt{3}}I_3$),
and spin-0 ($\frac{1}{\sqrt{2}}I_2$) and spin-1 ($\frac{1}{\sqrt{2}}\sigma_3$)
states, we find the relevant Fierz transformations to be
\begin{eqnarray}
d_1&=&4\pi\frac{1}{6}\left[4b_{8}+\frac{4}{3}a_{8}+3b_{1}+a_{1}\right]\; ,\nonumber\\
d_2&=&4\pi\frac{1}{6}\left[-\frac{4}{3}b_{8}+\frac{4}{3}a_{8}-b_{1}+a_{1}\right]\; ,\nonumber\\
d_3&=&4\pi\left[-\frac{3}{2}b_{8}-\frac{1}{6}a_{8}+3b_{1}+a_{1}\right]\; ,\nonumber\\
d_4&=&4\pi\left[\frac{1}{2}b_{8}-\frac{1}{6}a_{8}-b_{1}+a_{1}\right]\; .
\label{eq:d_ops}
\end{eqnarray}
Note that the factor of $4\pi$ is needed to re-express the contact operators
in terms of $\alpha_s^2$ rather than $\alpha_s g^2$.

From Eq. (\ref{eq:Sd}) we see that the energy of colour singlet
$\overline{Q}Q$ mesons receives a contributions proportional to
$d_1$ in the spin singlet case and to $d_2$ for the spin triplet case.
The hyperfine splitting is therefore proportional to $(d_1-d_2)$,
cf.~subsection \ref{sec:application}.\ref{subsec:hfs} for further discussion.
From Eq. (\ref{eq:d_ops}) we then find 
\BE
d_1-d_2 = \frac{32\pi}{9}b_8 + \frac{8\pi}{3}b_1\; ,
\EE
and so the coefficients $a_1$ and $a_8$ of the spin-independent operators
have cancelled out.
In what follows we therefore calculate the coefficients $b_8$ and $b_1$ of
the spin-dependent operators in Eq. (\ref{eq:Sab}).

Since there is no $\bs\cdot\bA$ coupling in NRQCD, it turns out that 1PR
diagrams cannot contribute to the spin-dependent non-derivative operators
and hence the operator combination which gives rise to the hyperfine
splitting can be matched using 1PI diagrams only.
In order to complete the matching of all four-fermion operators, including
the spin-independent contributions, we need to also calculate the 1PR
contributions from inserting the vacuum polarization onto a Coulomb gluon
exchange line, since this will generate four-fermion contact operators
to all orders in the derivatives.
It also turns out that the IR divergences in the spin-independent case
are more severe than for the spin-dependent case due in part to
these contributions from Coulomb exchange, which also require a careful
treatment of lattice artifacts multiplying logarithmic IR divergences.
We shall comment further on this observation below
and will present the general matching analysis for all non-derivative
four-fermion operators in a future paper.

\begin{figure}
\includegraphics[width=0.85\linewidth,clip=true]{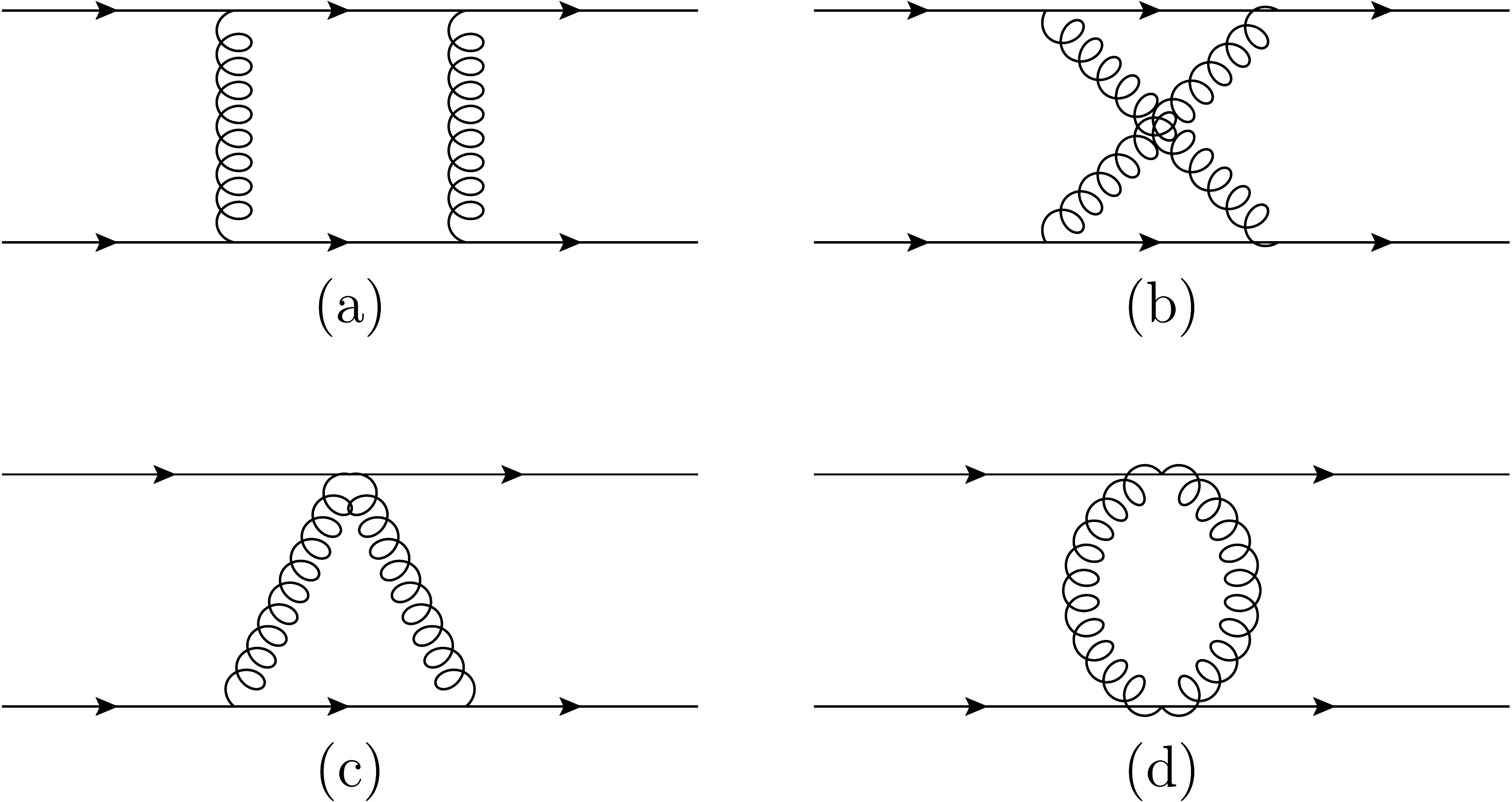}\;
\caption{Ladder Feynman diagrams.
         Referred to as (a) box, (b) crossbox, (c) triangle,
         and (d) diamond diagrams. Note that there are two diagrams with
         the triangle topology.}
\label{fig:boxdiags}
\end{figure}

As far as the spin-dependent interactions are concerned, the leading
contributions to the four-fermion interactions arise from the 1PI ladder
diagrams shown in Figure \ref{fig:boxdiags}.
The box (a) and crossbox (b) diagrams contribute in both continuum QCD and
NRQCD, whereas the triangle (c) and diamond (d) diagrams are specific to
NRQCD; again, let us emphasize that these NRQCD-specific diagrams
consist not only of lattice artifacts, but also contain continuum contributions
from vertices arising from the non-relativistic form of the NRQCD action.
Much as in the previous sections, by considering the 1PI diagrams
in both continuum QCD and lattice NRQCD, we can then determine the values
of the coefficients of the spin-dependent four-fermion operators in the
NRQCD action needed to account for radiative improvement corrections.

\begin{figure}[h]
\centering
\includegraphics[width=0.65\linewidth,clip=true]{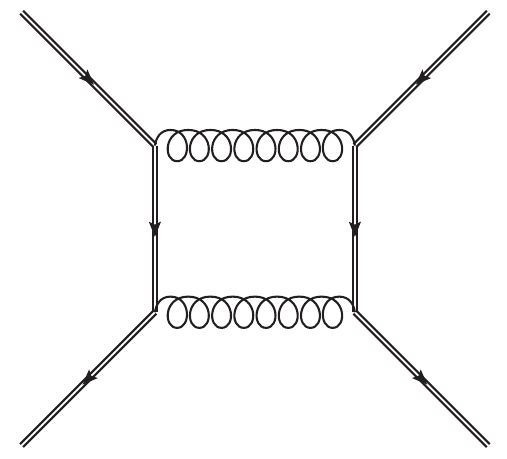}
\caption{Two-gluon annihilation diagram}
\label{fig:d1ann}
\end{figure}

In addition to the corrections due to $\overline{Q}Q \to \overline{Q}Q$
scattering calculated above, a further correction to $d_1$ is required
to account for the fact that the NRQCD formalism does not allow for the
creation or annihilation of $\overline{Q}Q$ pairs. Figure \ref{fig:d1ann}
shows the relevant continuum QCD diagram that is absent in NRQCD.
Labelle et al.~\cite{Labelle:1997}
give the amplitude for this process for QED and it is straightforward
to obtain the corresponding QCD result by including the correct colour factors
to give
\cite{Pineda:1998kj}
\begin{eqnarray}
d_1^{2-\gamma ann} = -\frac{2}{9}\left(2-2\ln{2}\right).
\end{eqnarray}

%%%%%%%%%%%%%%%%%%%%%%%%%%%%%%%%%%%%%%%%%%%%%%%%%%%%%%%%%%%%%%%%%%%%%%%%%%%%%%
\subsection{Calculation}
\label{subsec:4fcalc}

The absence of a $\overline{Q}Q$ vertex in NRQCD and the fact that
quarks and antiquarks have identical masses by the CPT theorem means
that the simplest implementation of antiquarks using the \textsc{HPsrc}
package is to treat them as just a different species of quark.
In fact, we can simply relate $QQ$-scattering diagrams
\begin{align}
&-\Gamma_{\psi^\dagger\psi\psi^\dagger\psi} = \\
&~~Z_{sym}^{\NR, (1)}\frac{\alpha_s g^2}{M^2}(\psi^\dagger_\alpha(iT_aiT_b)_{\alpha\delta}
  \psi_\delta)(\psi^\dagger_\beta(iT_aiT_b)_{\beta\gamma}\psi_\gamma) \nonumber\\
&~~+ Z_{asym}^{\NR,(1)}\frac{\alpha_s g^2}{M^2}(\psi^\dagger_
  \alpha(iT_biT_a)_{\alpha\delta}\psi_\delta)\psi^\dagger_\beta
(iT_aiT_b)_{\beta\gamma}\psi_\gamma)\nonumber\\
&~~+ Z_{sym-\sigma}^{\NR,(1)}\frac{\alpha_s g^2}{M^2}(\psi^\dagger_\alpha
i \sigma(iT_aiT_b)_{\alpha\delta}\psi_\delta)(\psi^\dagger_\beta
i\sigma (iT_aiT_b)_{\beta\gamma}\psi_\gamma)\nonumber\\
&~~+ Z_{asym-\sigma}^{\NR,(1)}\frac{\alpha_s g^2}{M^2}(\psi^\dagger_\alpha
i \sigma(iT_biT_a)_{\alpha\delta}\psi_\delta)(\psi^\dagger_\beta
i\sigma (iT_aiT_b)_{\beta\gamma}\psi_\gamma)\; ,\nonumber
\label{eq:Z4fops}
\end{align}
to the $\overline{Q}Q$ one by changing the representation of one of the
quarks appropriately. For NRQED, the antiparticle vertex can be obtained
by replacing $e\rightarrow-e$; the NRQCD analogue is taking
$(\sigma,T_a)\rightarrow(-\sigma^*,-T_a^T)$ to obtain%
\footnote{
We have used the convention that the anti-quark fields are in the
$\overline{\bf 2}$ representation, whereas some other authors put them in the
${\bf 2}$ representation. The two conventions can be related using the
charge conjugation matrix $C=i\sigma_2$. }
\begin{align}
&\Gamma_{\chi^\dagger\chi\psi^\dagger\psi} = \\
&~~-Z_{sym}^{\NR, (1)}\frac{\alpha_s g^2}{M^2}(\chi^\dagger_\alpha(T_bT_a)^T_{\alpha\delta}
\chi_\delta)(\psi^\dagger_\beta (T_aT_b)_{\beta\gamma}\psi_\gamma)
\nonumber\\
&~~- Z_{asym}^{\NR,(1)}\frac{\alpha_s g^2}{M^2}(\chi^\dagger_
\alpha(T_aT_b)^T_{\alpha\delta}\chi_\delta)(\psi^\dagger_\beta
(T_aT_b)_{\beta\gamma}\psi_\gamma)\nonumber\\
&~~- Z_{sym-\sigma}^{\NR,(1)}\frac{\alpha_s g^2}{M^2}(\chi^\dagger_\alpha
\sigma^*(T_bT_a)^T_{\alpha\delta}\chi_\delta)(\psi^\dagger_\beta\sigma
(T_aT_b)_{\beta\gamma}\psi_\gamma)\nonumber\\
&~~- Z_{asym-\sigma}^{\NR,(1)}\frac{\alpha_s g^2}{M^2}(\chi^\dagger_\alpha
\sigma^*(T_aT_b)^T_{\alpha\delta}\chi_\delta)(\psi^\dagger_\beta\sigma
(T_aT_b)_{\beta\gamma}\psi_\gamma)\; ,\nonumber
\end{align}
where the transpose of the individual $T_a$ elements has been pulled
outside and the antiquark fields have been relabelled to $\chi$
to agree with convention.
We have kept all indices explicit to highlight how the creation
and annihilation operators are directly replaced.

Dividing out the colour factors and isolating the correct spin
structure on both fermion lines passing through the diagram, we can
calculate the one-loop coefficients, $Z_{sym}^{\NR,(1)}$ etc., directly
in the \textsc{HPsrc} code. The factors of $i$ in Eq. (\ref{eq:Z4fops})come from the convention
that we work with anti-hermitian generators when evaluating the colour
factors of diagrams.
This is convenient since then the commutation relations have real
structure constants. Note that by construction the box diagram (a) contributes
only to the symmetric colour combination, whereas the crossbox diagram (b)
contributes only to the antisymmetric colour combination.
The triangle (c) and diamond (d) diagrams contribute to both colour
combinations since they contain two-gluon vertices carrying colour
factors $\{T_a,T_b\}$.

Although the momentum exchange acts as an IR regulator in these diagrams,
we set all three-momenta to zero with the external quarks on-shell and use
a gluon mass $\mu$ as the IR regulator using a St\"uckelberg mass term.
This is possible as discussed above, and helps the evaluation by
significantly simplifying the pole structure of the integrand: all diagrams
have a simple pinch at $k_0=\pm \mu$, and the calculation can be performed
without needing to implement contour shifts in the complex energy plane.
As in the case of the BFG vertices, the absence of UV divergences allows
us to employ different UV regulators in the QCD and NRQCD calculation.

Radiative improvement is then achieved by adding four-fermion operators to the
lattice NRQCD action with coefficients tuned such that one-loop calculations
of $\overline{Q}Q\to\overline{Q}Q$ scattering in NRQCD give the correct
continuum QCD result. For example, we add terms such as
\begin{equation}
\Gamma = -Z_{sym}^{(1)}\frac{\alpha_s g^2}{M^2}(\chi^\dagger_\alpha(T_bT_a)^T_{\alpha\delta}\chi_\delta)
(\psi^\dagger_\beta(T_aT_b)_{\beta\gamma}\psi_\gamma)\;,
\end{equation}
where $Z_{sym}^{(1)}=Z_{sym}^{QCD,(1)}-Z_{sym}^{\NR,(1)}$ is the 
difference between the diagrammatic coefficients calculated in continuum QCD
and NRQCD, respectively. The radiative improvement can then be implemented
with the coefficients given by
\begin{eqnarray}
a_{8} 
&=&-\frac{1}{4}\{(3+\frac{5}{3})Z_{sym}^{(1)}+(-3+\frac{5}{3})Z_{asym}^{(1)}\}\; 
,\label{eq:a8}\\
a_{1} &=&-\frac{2}{9}\{Z_{sym}^{(1)}+Z_{asym}^{(1)}\}\; ,\label{eq:a1}\\
b_{8} 
&=&-\frac{1}{4}\{(3+\frac{5}{3})Z_{sym-\sigma}^{(1)}+(-3+\frac{5}{3})Z_{asym-\sigma}^{(1)}\}\; 
,\label{eq:b8}\\
b_{1} &=&-\frac{2}{9}\{Z_{sym-\sigma}^{(1)}+Z_{asym-\sigma}^{(1)}\}\; . \label{eq:b1}
\end{eqnarray}

The analytical calculation of the continuum QCD diagrams is relatively 
simple with our choice of IR regulator. The results are
\begin{eqnarray}
Z_{sym}^{QCD,(1)}&=&-\frac{7}{12\pi}+\frac{1}{4\pi}\log{(\mu/M)}+
                   \frac{5M}{16\mu}-\frac{M^2}{2\pi\mu^2}+\frac{M^3}{2\mu^3}\; ,\nonumber\\
Z_{asym}^{QCD,(1)}&=&-\frac{3}{4\pi}\log{(\mu/M)}-\frac{3M}{8\mu}+\frac{M^2}{2\pi\mu^2}\; ,\\\label{eq:a_coeff}
Z_{sym-\sigma}^{QCD,(1)}&=&-\frac{1}{4\pi}+\frac{1}{4\pi}\log{(\mu/M)}+\frac{M}{6\mu}\; ,\nonumber\\
Z_{sym-\sigma}^{QCD,(1)}&=&-\frac{1}{4\pi}\log{(\mu/M)}\; .\label{eq:b_coeff}
\end{eqnarray}

\begin{figure}[t]
\centering
\includegraphics[width=0.9\linewidth,clip=true]{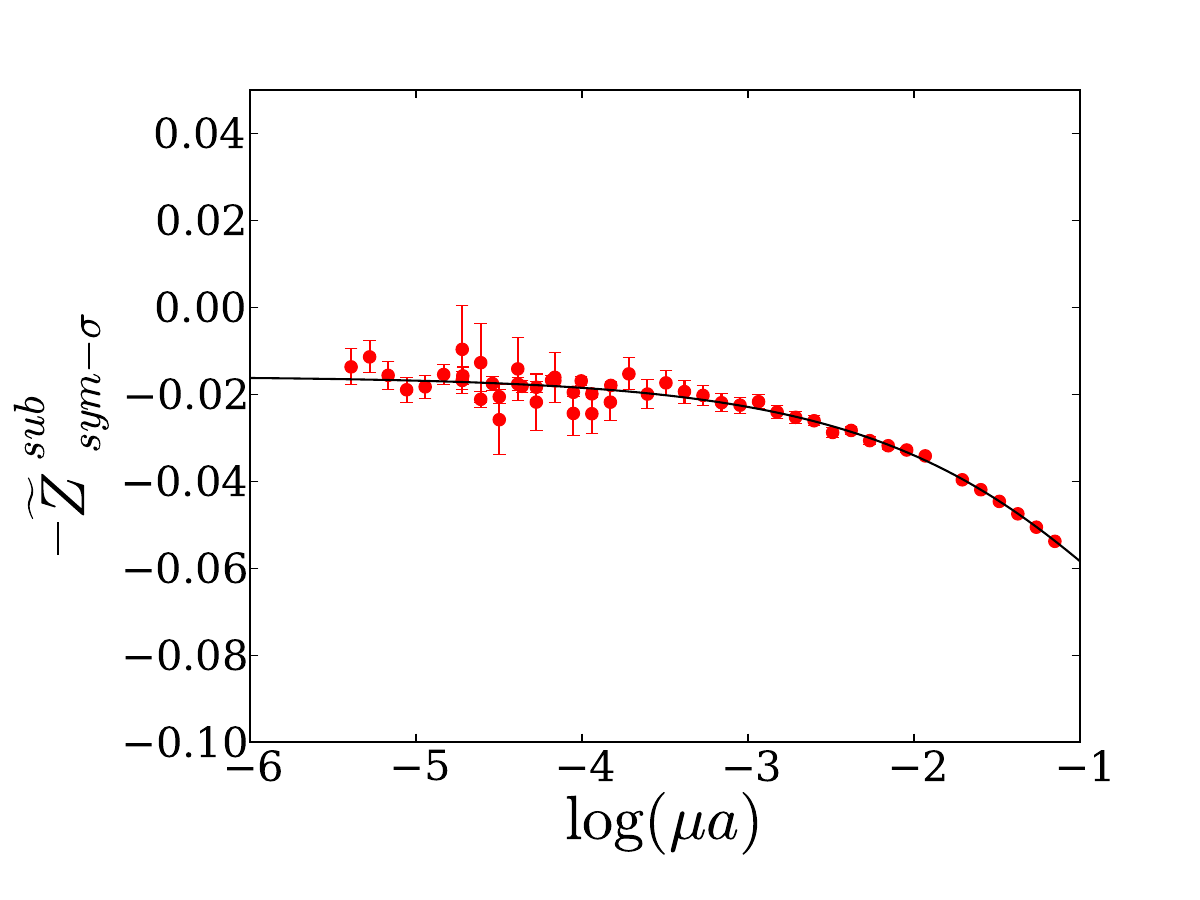}
\caption{Sample fit for $\widetilde Z^{sub}_{sym-\sigma}(\mu)$ for $Ma=1.95$. 
         We obtain $-0.0159(17)-0.145(15)\mu a +0.08(3)(\mu a)^2$
         with $\chi^2/d.o.f.=0.09$.}
\label{fig:4fsubplot3}
\end{figure}

\begin{figure}[t]
\centering
\includegraphics[width=0.90\linewidth,clip=true]{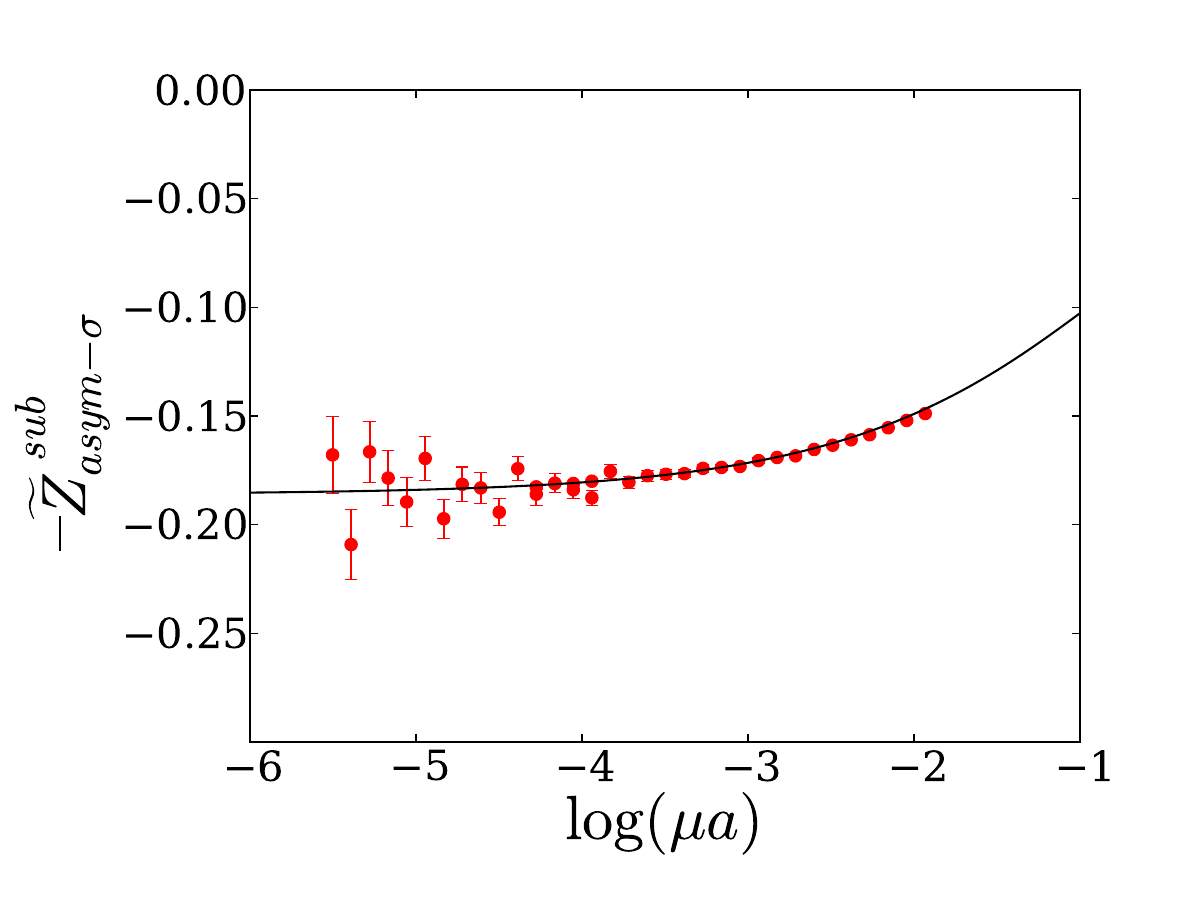}
\caption{Sample fit for $\widetilde Z_{asym-\sigma}^{sub}(\mu)$ for $Ma=1.95$.
         We obtain $-0.186(3)+0.3(1)\mu a-0.2(3)(\mu a)^2$ with $\chi^2/d.o.f.=0.08$.}
\label{fig:4fsubplot4}
\end{figure}

The NRQCD diagrams are computed numerically using the \textsc{HPsrc} 
package. Since higher order IR divergences dominate numerically, suitable
subtraction functions must be chosen to pointwise cancel the divergent
integrands:
\begin{widetext}
\begin{align}
I^{\rm sub}_{\rm sym}(\mu) &= - \int \frac{d^4 k}{(2\pi)^4} \frac{4\pi M^2(1+{\bf k}^2/4M^2)}{(k^2+\mu^2)^2(k_0^2+{\bf k}^4/4M^2)} 
  &&= \tilde{Z}^{\rm sub}_{\rm sym}(\mu) -\frac{1}{\pi}\log(\mu a)
     -\frac{5M}{16\mu} +\frac{M^2}{2\pi\mu^2} -\frac{M^3}{2\mu^3} \\
I^{\rm sub}_{\rm asym}(\mu) &= - \int \frac{d^4 k}{(2\pi)^4} \frac{4\pi M^2}{(k^2+\mu^2)^2(ik_0+{\bf k}^2/2M)^2} 
  &&= \tilde{Z}^{\rm sub}_{\rm asym}(\mu) +\frac{15}{8\pi}\log(\mu a)
     +\frac{3M}{8\mu} -\frac{M^2}{2\pi\mu^2} \\
I^{\rm sub}_{\rm sym-\sigma}(\mu) &= - \int \frac{d^4 k}{(2\pi)^4} \frac{4\pi}{(k^2+\mu^2)^2}\left(\frac{{\bf k}^2/3}{k_0^2+{\bf k}^4/4M^2}+\frac{1}{2}\right) 
  &&= \tilde{Z}^{\rm sub}_{\rm sym-\sigma}(\mu) -\frac{1}{4\pi}\log(\mu a)
     -\frac{M}{6\mu} \\
I^{\rm sub}_{\rm asym-\sigma}(\mu) &=- \int \frac{d^4 k}{(2\pi)^4} \frac{4\pi}{(k^2+\mu^2)^2}\left(\frac{{\bf k}^2/3}{(ik_0+{\bf k}^2/2M)^2}-\frac{1}{2}\right) 
  &&= \tilde{Z}^{\rm sub}_{\rm asym-\sigma}(\mu) +\frac{1}{4\pi}\log(\mu a)
\end{align}
\end{widetext}
These subtraction functions are continuum-like in that they contain no
lattice artifact IR divergences. Evaluating $I^{sub}$ numerically and
fitting (with the coefficients of the divergences constrained to agree
with the analytical results, and a polynomial in $\mu a$ added) allows
us to determine $\widetilde Z^{sub}$. As the integrands are easily evaluated,
a large number of points can be sampled. 
The high-order divergences cause the integrand to become extremely large 
around $k\sim 0$, which can mean that for small enough $k$ even
double-precision variables can overflow, thus returning \texttt{NaN} for 
the value of the integrand. To avoid this, we add a statement that will 
set the integrand to zero in a small neighbourhood of the origin.
We have to use a sufficient number of VEGAS sampling points in order to
resolve this cut around the origin exactly. In
Figures \eqref{fig:4fsubplot3}-\eqref{fig:4fsubplot4} we plot sample fits
of these subtraction functions.

For each colour ordering, the sum of all diagrams is then calculated together  
with the appropriate subtraction function. In this way all IR divergences 
except, in some cases, the logarithmic IR divergence are cancelled, regardless 
of the lattice artifacts, allowing a more constrained fit. For the
spin-dependent contribution we calculate
\small
\begin{eqnarray}
Z_{sym-\sigma}^{\NR,(1)}+I^{sub}_{sym-\sigma}(\mu)-\widetilde 
Z^{sub}_{sym-\sigma}(0)&=&\widetilde Z_{sym-\sigma}^{\NR,(1)}\; ,\\
Z_{asym-\sigma}^{\NR,(1)}+I^{sub}_{asym-\sigma}(\mu)-\widetilde 
Z^{sub}_{asym-\sigma}(0)&=&\widetilde Z_{asym-\sigma}^{\NR,(1)}\; .
\end{eqnarray}
\normalsize
We evaluate the diagrams for several gluon masses in the range
$10^{-3}<\mu^2<10^{-2}$, which is chosen such that any lattice artifacts
(appearing as polynomials in $\mu a$) are negligible, yet at the same time
large enough such that we do not run foul of any hard-wired numerical
tolerances in the \textsc{HPsrc} package.

%%%%%%%%%%%%%%%%%%%%%%%%%%%%%%%%%%%%%%%%%%%%%%%%%%%%%%%%%%%%%%%%%%%%%%%%%%%%%%
\subsection{Results}
\label{subsec:4fres}

We then have
\begin{eqnarray}
Z_{sym-\sigma}^{(1)}&=&-\widetilde 
Z_{sym-\sigma}^{\NR,(1)}-\frac{1}{4\pi}-\frac{1}{4\pi}\log{Ma}\;,\nonumber\\
Z_{asym-\sigma}^{(1)}&=&-\widetilde 
Z_{asym-\sigma}^{\NR,(1)}+\frac{1}{4\pi}\log{Ma}\;.\nonumber
\end{eqnarray}
Using these equations, the numerical results for the radiative improvement 
coefficients $b_1$ and $b_8$ for the spin-dependent operators are calculated 
from Eq. (\ref{eq:b8}) and (\ref{eq:b1}). Results for various NRQCD actions
are given in Tables \eqref{tab:4fspindep1} to \eqref{tab:4fspindep3}.
Note that the coefficients have a sizeable dependence on $Ma$.
For the full spin $v^6$ NRQCD action, this dependence is less pronounced.  

As noted at the beginning of this section, we will report on the calculation of
the radiative improvement coefficients $a_1$ and $a_8$ for the
spin-independent operators in Eq. (\ref{eq:Sab}) in a future paper.

\begin{table}
\centering
\begin{tabular}{|c|ccccc|}
\hline
$Ma$ & 1.95 &~~& 2.8 &~~& 4.0 \\\hline\hline
$b_1$ & 0.0037(3) && -0.0201(4)   && -0.0490(5) \\
$b_8$ & 0.0893(8) &&  0.0183(15) && -0.0832(26) \\
\hline
\end{tabular}
\caption{Renormalization parameters for spin-dependent four-fermion operators,
         full $v^4$ NRQCD $n=2$}
\label{tab:4fspindep1}
\end{table}

\begin{table}
\centering
\begin{tabular}{|c|ccccc|}
\hline
$Ma$ & 1.9 &~~& 2.65 &~~& 3.4 \\\hline\hline
$b_1$ & 0.0075(1) && -0.0148(6) && -0.0341(2) \\
$b_8$ & 0.0997(5) &&  0.0353(1) && -0.0290(8) \\
\hline
\end{tabular}
\caption{Renormalization parameters for spin-dependent four-fermion operators,
         full $v^4$ NRQCD $n=4$}
\label{tab:4fspindep2}
\end{table}

\begin{table}
\centering
\begin{tabular}{|c|ccccc|}
\hline
$Ma$ & 1.9 &~~& 2.65 &~~& 3.4 \\\hline\hline
$b_1$ & -0.0223(2) && -0.03599(2) && -0.0504(2) \\
$b_8$ & -0.0280(9) && -0.0624(7)  && -0.1071(9) \\
\hline
\end{tabular}
\caption{Renormalization parameters for spin-dependent four-fermion operators,
         full-spin $v^6$ NRQCD $n=4$}
\label{tab:4fspindep3}
\end{table}

%%%%%%%%%%%%%%%%%%%%%%%%%%%%%%%%%%%%%%%%%%%%%%%%%%%%%%%%%%%%%%%%%%%%%%%%%%%%%%%%
%%%%%%%%%%%%%%%%%%%%%%%%%%%%%%%%%%%%%%%%%%%%%%%%%%%%%%%%%%%%%%%%%%%%%%%%%%%%%%%%
\section{\texorpdfstring{Application to $\Upsilon$ and $\eta_b$ spectrum}%
                        {Application to bottomonium spectrum}}
%%%%%%%%%%%%%%%%%%%%%%%%%%%%%%%%%%%%%%%%%%%%%%%%%%%%%%%%%%%%%%%%%%%%%%%%%%%%%%%%
%%%%%%%%%%%%%%%%%%%%%%%%%%%%%%%%%%%%%%%%%%%%%%%%%%%%%%%%%%%%%%%%%%%%%%%%%%%%%%%%
\label{sec:application}

Having derived the radiative corrections to the $\sigma\cdot B$,
Darwin, and spin-dependent four-fermion operators in the NRQCD action,
we proceed to analyse the effects of these corrections on the spectrum of
mesons containing b-quarks. Depending on whether the operators involved
are spin-dependent or not, we can distinguish between
changes to the hyperfine splitting and overall shifts of the ground state mass.
While obviously no substitute for simulations including the radiatively
corrected coefficients in the NRQCD action, estimates for these effects
are important because they give a clear indication about the expected
magnitude of the corrections to the spectrum of S-wave $b\bar b$ states. 
In what follows the quark mass, $M$, is identified with the mass, $M_b$, of the $b$-quark
and defined on the lattice as $a^{-1}(aM_b)$.

%%%%%%%%%%%%%%%%%%%%%%%%%%%%%%%%%%%%%%%%%%%%%%%%%%%%%%%%%%%%%%%%%%%%%%%%%%%%%%
\subsection{Hyperfine Splitting}
\label{subsec:hfs}

\begin{figure}[t]
\centering
\includegraphics[width=0.65\linewidth,clip=true]{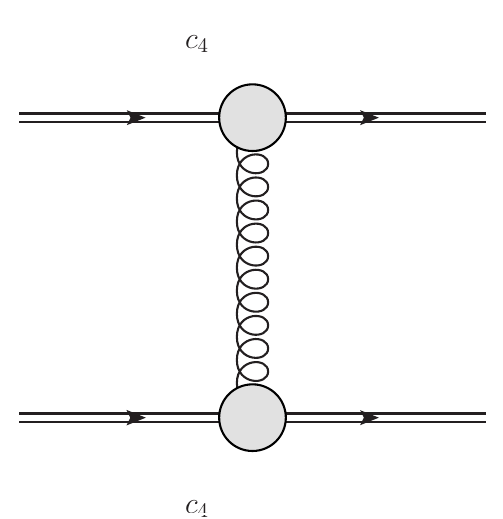}
\caption{The tree-level diagram contributing to heavy-heavy meson
         hyperfine splitting: a spatial gluon is exchanged between
         two vertices involving the chromomagnetic operator.}
\label{fig:spinexchange}
\end{figure}

The leading contribution to the hyperfine splitting can be estimated
from a perturbative picture involving the exchange of a single gluon
between two vertices involving the chromomagnetic operator
(cf. Figure \ref{fig:spinexchange}); this process shifts the energy
of a colour-singlet meson state
$|M\ra=\frac{1}{\sqrt{3}}|r\bar{r}+b\bar{b}+g\bar{g}\ra|S\ra$ by
\begin{eqnarray}
\Delta E&\approx&\frac{c_4^2g^2}{4M^2}\la M|(\sigma\times iq T_a)(\sigma^*\times iq T_a^T)|M\ra 
/q^2
\nn\\
&=& \frac{c_4^2g^2}{4M^2}\la S|(\sigma\times iq)(\sigma^*\times iq)|S\ra\frac{1}{3}Tr(T_aTa)/q^2
\nn\\
&=&\frac{-2c_4^2g^2}{9M^2}\la S|\sigma.\sigma^*|S\ra\; .
\end{eqnarray}
For $|\eta_b\ra=\frac{1}{\sqrt{2}}|\uparrow\tilde{\uparrow}+\downarrow\tilde{\downarrow}\ra$ (S=0) and 
$|\Upsilon{}>=|\uparrow\tilde{\downarrow}\ra$ (S=1),
we have
$\la\eta_b|\sigma.\sigma^*|\eta_b\ra-\la\Upsilon{}|\sigma.\sigma^*|\Upsilon{}\ra=3-(-1)=4$, and hence 
the hyperfine splitting is approximately
\begin{equation}
M_{\eta_b}-M_{\Upsilon}\approx\frac{-8c_4^2g^2}{9M^2}|\psi(0)|^2 \;,
\end{equation}
where $\psi(0)$ is the meson wavefunction at the origin, assumed to be the same for the $\eta_b$ and
$\Upsilon{}$ states. We see that the $\eta_b$ lies below the $\Upsilon{}$, as expected.

Nonperturbative results indicate that the hyperfine splitting
in heavy quarkonium is indeed approximately proportional to $c_4^2$
\cite{Meinel10}.
Given that the origin of these nonperturbative results can be understood
from a tree-level estimate,
we can proceed to examine the effect of the four-fermion operators in
the same manner.
This is straightforward when working in the basis of operators
given by Eq. (\ref{eq:Sd}), where that we get
\begin{equation}
M_{\eta_b}-M_{\Upsilon}\approx \frac{9(d_1-d_2)}{2}\alpha_s^2\frac{4}{3M^2}|\psi(0)|^2
\end{equation}
for the four-fermion contribution to the hyperfine splitting, which can
be re-expressed in term of the coefficients $a_i$, $b_i$ using
Eq. (\ref{eq:d_ops}).

The one-loop correction to the hyperfine splitting is then estimated to be the tree-level value
multiplied by the one-loop correction factor
\begin{widetext}
\begin{equation}
1~+~\alpha_V(q^*)\left(2\,c_4^{(1)}-\frac{9}{8}\left(\frac{16}{3}b_{8}+4b_{1}\right)+\frac{3}{8\pi}
\left(2-2\ln{2}\right)\right)\;. 
\label{eq:hfs}
\end{equation}
\end{widetext}

\begin{table*}[t]
\centering
\small
\begin{tabular}{|c c c| c c c |c |c|}
\hline
M & $\beta$ & $\alpha_V$ & $c_4$ correction & box correction & total & old hfs & new 
hfs\\ [0.5ex]
\hline
1.95 & 7.09 & 0.216 & +31.4(3)$\%$ & -10.4(1)$\%$  & +21.0(3)$\%$ & 56(2)& 68(3)(5)(6) \\
2.8 & 6.76 & 0.249 & +39.8(3)$\%$  & +1.3(2)$\%$ & +41.1(4)$\%$ & 50(2)& 71(3)(6)(5)  \\
4.0 & 6.458 & 0.293 & +49.3(3)$\%$  & +23.2(3)$\%$ & +72.5(4)$\%$ & 41(2)& 71(3)(7)(4) \\[1ex]
\hline
\end{tabular}
\normalsize
\caption{Estimates of the corrections to the bottomonium hyperfine splitting
         results of
         \cite{PhysRevD.72.094507}
         arising from the radiative improvement of the $n=2$ full $v^4$ NRQCD
         action. The errors given in the last column are statistical,
         $O(\alpha_s^2)$, and relativistic corrections, in that order.}
\label{tab:n2hfs}
\end{table*}

\begin{figure}[t]
\centering
\includegraphics[width=0.95\linewidth,clip=true]{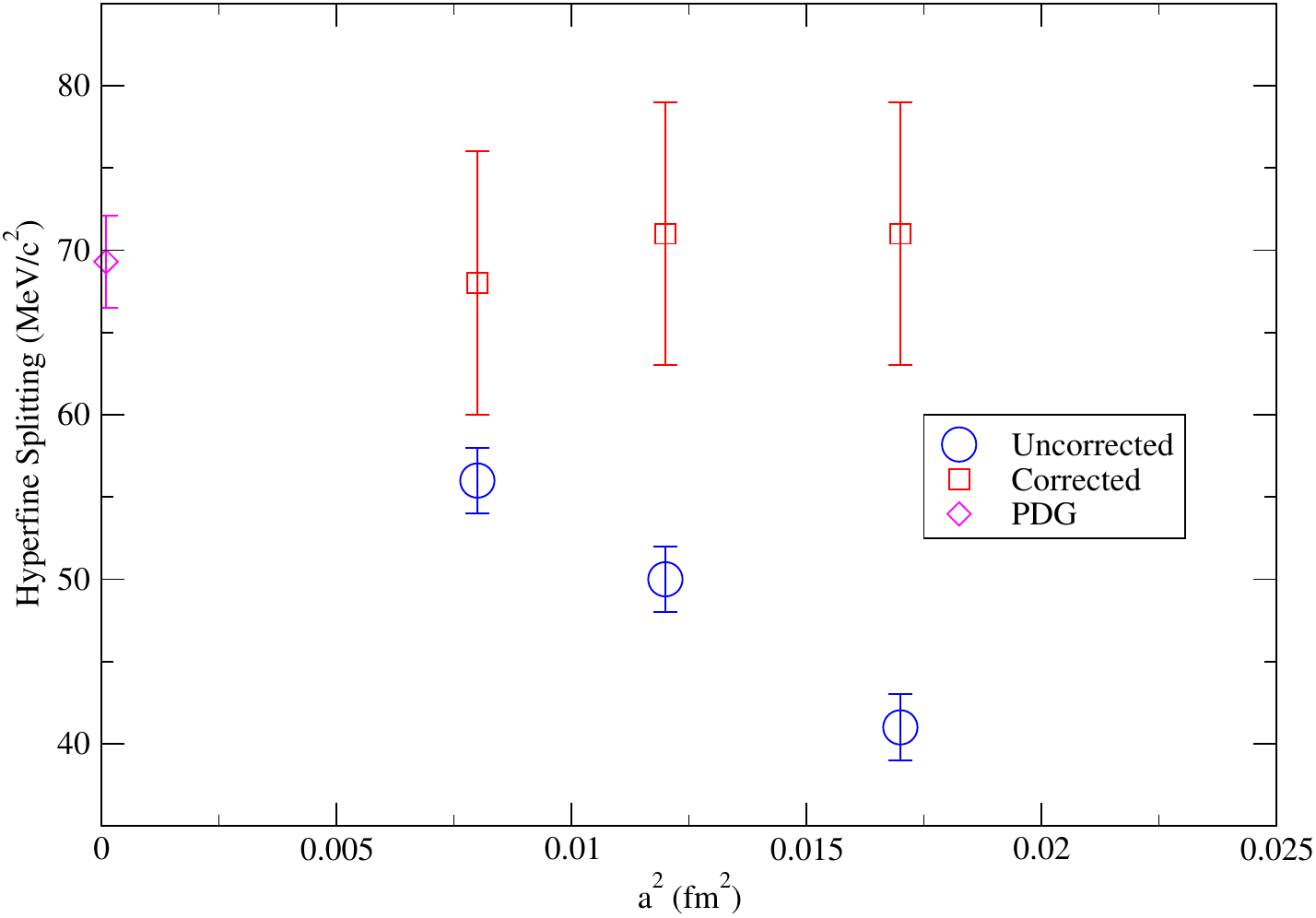}
\caption{Comparison of the corrected and uncorrected bottomonium hyperfine
         splitting results for $n=2$ full $v^2$ NRQCD
         \cite{PhysRevD.72.094507}. 
         Note that for the corrected results the total error including estimates
         of the effects of $O(\alpha_s^2)$ contributions and $O(v^6)$ terms
         omitted in the simulation is displayed,
         whereas the uncorrected results are shown with statistical errors
         only, since their $O(\alpha_s)$ errors would be too large to show
         on this scale. The PDG data point is taken from
         \cite{PDG:2010}.
}
\label{fig:n2hfsplot}
\end{figure}

In Table \ref{tab:n2hfs} we give our estimates for the corrections that
need to be applied to the hyperfine splitting as measured in full
$v^4$ NRQCD $n=2$ without perturbative improvements. Applying our estimated
correction {\em post hoc} to the data presented by the HPQCD collaboration in
\cite{PhysRevD.72.094507},
we find that the corrections to the chromomagnetic operator and the inclusion
of the spin-dependent four-fermion terms work to increase the hyperfine
splitting, pushing it closer to the experimental value of 69.3(2.8)MeV
\cite{PDG:2010}.
We note that the corrections from the four fermion operators further reduce
the lattice spacing dependence, as can be seen from the results plotted in Figure 
\ref{fig:n2hfsplot}.  It is important to note that for the $O(v^4)$ NRQCD action 
the errors on the corrected results include an estimate of the effect of the omitted 
$O(v^6)$ term shown in Eq. (\ref{eq:v6}), and that these errors are preliminary in the sense
that future simulations will include these terms explicitly, using the results of this paper, 
and removing the need for a {\em post hoc} correction.

\begin{table}[t]
\centering
\begin{tabular}{|cc|cc|c|c|}
\hline
&&\multicolumn{2}{|c|}{~hfs (MeV)~}&Correction&hfs (MeV)\\
$Ma$  & $\alpha_V(q^*)$& $c_4=1$ & improved $c_4$&4-fermion&corrected\\\hline\hline
1.9   & 0.225        & 56.1(1) &72.1(1)& -12.6(1)\% &65.0(1)(2.8)(5.6) \\
2.65  & 0.253        & 50.5(1) &69.8(1)& -1.8(1)\% &68.9(1)(3.2)(5.0) \\
3.4   & 0.275        & 45.6(1) &65.6(1)& +11.0(1)\% &70.6(1)(3.4)(4.6) \\
\hline
\end{tabular}
\caption{The corrections to the bottomonium hyperfine splitting
         arising from the radiative improvement of the $n=4$ full $v^4$ NRQCD
         action, as found in
         \cite{Dowdall:2011wh}.
         Only the four-fermion contributions are {\em post hoc} estimates.
         The errors given in the last column the errors are statistical,
         $O(\alpha_s^2)$, and relativistic corrections, in that order.}
\label{tab:n4hfs}
\end{table}
\begin{figure}[t]
\centering
\includegraphics[width=0.95\linewidth,clip=true]{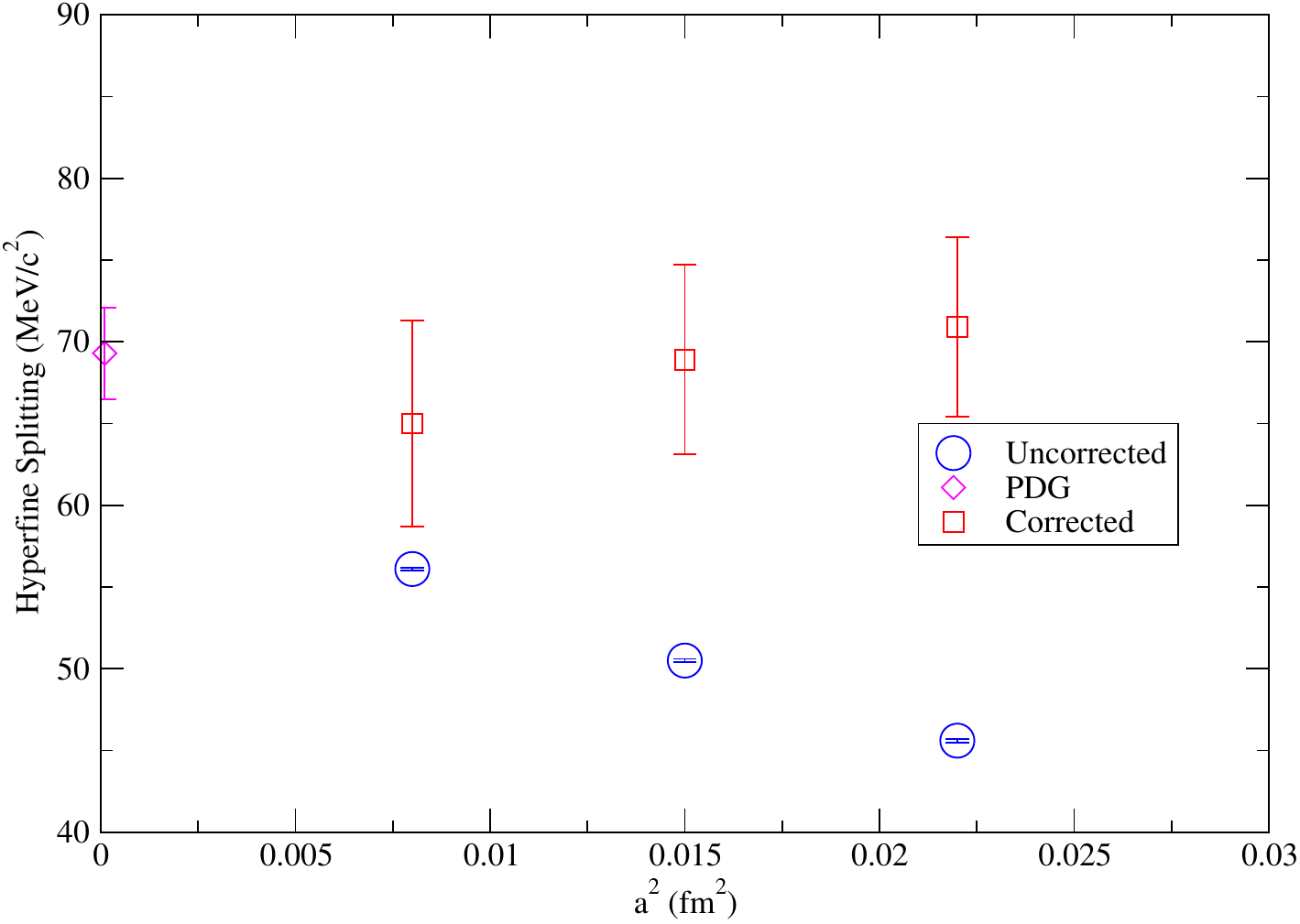}
\caption{Comparison of the corrected and uncorrected bottomonium hyperfine
         splitting results for $n=4$ full $v^4$ NRQCD
         \cite{Dowdall:2011wh}. 
         Note that for the corrected results the total error including estimates
         of the effects of $O(\alpha_s^2)$ contributions and $O(v^6)$ terms
         omitted in the simulation is displayed,
         whereas the uncorrected results are shown with statistical errors only,
         since their $O(\alpha_s)$ errors would be too large to show on this scale.
         The PDG data point is taken from
         \cite{PDG:2010}.
}
\label{fig:n4hfsplot}
\end{figure}

In Table \ref{tab:n4hfs} and Figure \ref{fig:n4hfsplot}
we present the corrections to the full $v^4$ NRQCD
$n=4$ hyperfine splitting as measured in
\cite{Dowdall:2011wh},
where the perturbatively corrected coefficient $c_4$ has been included
in the simulation, and only the four-fermion operator corrections
need to be applied by hand. As in $n=2$ case, the corrected chromomagnetic
operator acts to increase the hyperfine splitting, and
the four-fermion corrections reduce the lattice spacing dependence
(which however is less severe from the outset when comparing to the $n=2$ case).

\begin{table}[t]
\centering
\begin{tabular}{|c c| c c|}
\hline
M & $\alpha_V$ & $c_4$ correction & box correction\\ [0.5ex]
\hline\hline
1.9  & 0.225  & +37(4)$\%$  & +7.7(1)$\%$  \\
2.65 & 0.253  & +40(4)$\%$  & +15.4(1)$\%$   \\
3.4  & 0.275  & +26(4)$\%$  & +25.9(1)$\%$  \\ [1ex]
\hline
\end{tabular}
\caption{Corrections to the bottomonium hyperfine splitting arising from
         the radiative improvement of the $n=$ full NRQCD action including
         spin-dependent terms at order $v^6$.}
\label{tab:n4v6hfs}
\end{table}

\begin{figure}[t]
\centering
\includegraphics[width=0.65\linewidth,clip=true]{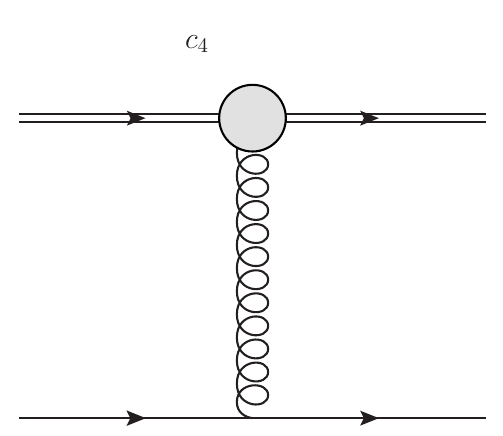}
\caption{The tree-level diagram contributing to the hyperfine splitting
         in heavy-light mesons: a spatial gluon is exchanged between two
         vertices, the heavy-quark one of which involves the chromomagnetic
         operator.}
\label{fig:heavylightspinexchange}
\end{figure}

Finally, in Table \ref{tab:n4v6hfs} we estimate the relative corrections 
from the corrected coefficients to full $n=4$ NRQCD results when including
the spin-dependent $v^6$ terms in the action.
While the corrections arising from the improvement of the chromomagnetic term
are similar to the case of full $n=4$ $v^4$ NRQCD, the corrections coming
from the four-fermion operators are significantly larger.
While one might naively expect that the effect of radiative corrections
ought to decrease as higher-order terms are added to the action,
it has been shown in
\cite{Meinel10}
that the inclusion of $O(v^6)$ terms (albeit with a different gauge action)
leads to a decrease of the (tree-level) hyperfine splitting. It appears that the larger
corrections from the four-fermion operators would compensate for this effect.   

The radiative corrections to the chromomagnetic operator (but not the
four-fermion operators) will also affect the hyperfine splitting in
heavy-light mesons through the leading perturbative contribution shown in
Figure \ref{fig:heavylightspinexchange}, giving a hyperfine splitting 
approximately proportional to $c_4$.
Besides the absence of the four-fermion terms, heavy-light systems also
benefit from much smaller $v^6$ corrections, which makes them a particularly
suitable test case for assessing the efficiency of including the radiative
corrections. In
\cite{Dowdall:2012ab},
the NRQCD action with radiatively improved coefficients was used for the first
time for the heavy-light $B$-meson system. The results for the hyperfine
splittings  for the $B$ and $B_s$ mesons show little $a^2$ dependence, and
even for coarse lattices are in very good agreement with experimental data;
this gives a strong check on the correctness and usefulness of the radiative
corrections and lends credibility to the prediction (rather than postdiction) of
$M_{B_c^*}-M_{B_c}=54(3)$~MeV
\cite{Dowdall:2012ab}
incorporating the effects of radiatively improving $c_4$.

%%%%%%%%%%%%%%%%%%%%%%%%%%%%%%%%%%%%%%%%%%%%%%%%%%%%%%%%%%%%%%%%%%%%%%%%%%%%%%
\subsection{Mass shift}
\label{subsec:shift}

\begin{figure}[b]
\centering
\includegraphics[width=0.65\linewidth,clip=true]{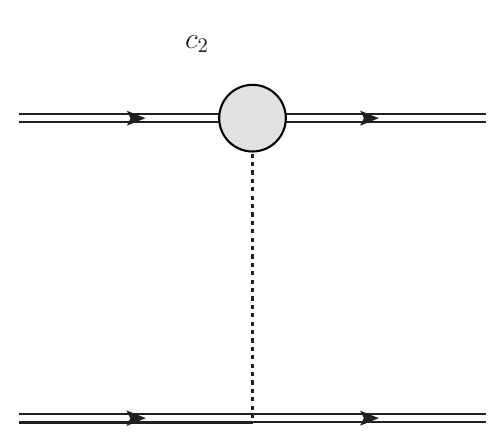}
\caption{The tree-level diagram contributing to the ground state
         mass shift: a temporal gluon is exchanged between two vertices,
         one of which involves the Darwin operator.}
\label{fig:exchange}
\end{figure}

The leading spin-independent perturbative correction to the energy
of a meson state is given by the single-gluon exchange involving the
Darwin term at one of the vertices, as shown in Figure \ref{fig:exchange}.
This gives the energy shift from the corrected $c_2$ coefficient as
\begin{eqnarray}
\Delta E&\approx&\frac{2c_2g^2}{8M^2}\la M|(q^2T_a)(T_a^T)|M\ra /q^2 \\
&=&\frac{-c_2g^2}{3M^2}|\psi(0)|^2.
\label{eq:shift1}
\end{eqnarray}

The corresponding contribution from the four-fermion operators gives
\begin{eqnarray}
\Delta E&\approx&\frac{9d_1}{2}\alpha_s^2\frac{4}{3M^2}|\psi(0)|^2\;.
\label{eq:shift2}
\end{eqnarray}

While the hyperfine splitting between the $\eta_b$ and $\Upsilon$ states
will lead to higher-order corrections to their wavefunctions at the origin,
to leading order these can be taken to be identical, here denoted $\psi(0)$, giving 
the same shift in mass for both states.

Again, there is no real substitute for including the four-fermion operators
in numerical simulations; however, we can attempt to estimate the size of
their overall effect by determining an effective value for $c_2$ using
the tree-level approximations given above.

Combining Eqs. (\ref{eq:shift1}) and (\ref{eq:shift2}), we find
\begin{widetext}
\begin{eqnarray}
\Delta E&\approx&\frac{-g^2}{3M^2}\left(1+\alpha_s(c_2^{(1)}-12b_8-4a_8-9b_1-3a_1+\frac{1}{\pi}
\left(2-2\ln{2}\right))\right)|\psi(0)|^2\; ,
\label{eq:DMeta_b}
\end{eqnarray}
\end{widetext}
giving an effective value of
$c_2^{eff}=1+\alpha_s(c_2^{(1)}-12b_8-4a_8-9b_1-3a_1+\frac{1}{\pi}\left(2-2\ln{2}\right))$. 
From this expression it is clear that we cannot consider the correction
due to the Darwin operator in isolation, but must include the effects of the
spin-independent four-fermion operators which we have not computed in this paper
and whose calculation will be presented in a forthcoming paper.

%%%%%%%%%%%%%%%%%%%%%%%%%%%%%%%%%%%%%%%%%%%%%%%%%%%%%%%%%%%%%%%%%%%%%%%%%%%%%%%%
%%%%%%%%%%%%%%%%%%%%%%%%%%%%%%%%%%%%%%%%%%%%%%%%%%%%%%%%%%%%%%%%%%%%%%%%%%%%%%%%
\section{Conclusions}
%%%%%%%%%%%%%%%%%%%%%%%%%%%%%%%%%%%%%%%%%%%%%%%%%%%%%%%%%%%%%%%%%%%%%%%%%%%%%%%%
%%%%%%%%%%%%%%%%%%%%%%%%%%%%%%%%%%%%%%%%%%%%%%%%%%%%%%%%%%%%%%%%%%%%%%%%%%%%%%%%
\label{sec:conclusions}

In this paper, we have applied the BF method to lattice NRQCD and have 
computed the one-loop radiative correction to the coefficient, $c_4$, of the 
$\vec{\sigma}\cdot\vec{B}$ and the one-loop radiative contribution to the 
coefficients, $d_1$ and $d_2$ of the four-fermion contact operators that affect 
the hyperfine structure of heavy quark mesons. The gauge independence of our 
calculation was explicitly checked by carrying out both relativistic and 
non-relativistic calculations in the lattice theory. This is possible because in 
BFG all calculations are UV finite.

Our results are summarized in
Tables \ref{tab:n2hfs}, \ref{tab:n4hfs}, and \ref{tab:n4v6hfs},
and in Eqns. (\ref{eq:c4}), (\ref{eq:hfs}) and (\ref{eq:DMeta_b}).
In particular, in eqn. (\ref{eq:c4}) there is a negative correction to $c_4$
due the the effect of the continuum  logarithmic IR divergence.
However, it turns out that the constant terms more than cancel this effect
and the correction to $c_4$ is positive.

Whilst no substitute for including these corrections in a simulation,
our estimate for the correction to the $\Upsilon - \eta_b$ hyperfine splitting
measured by
Gray et al.~\cite{PhysRevD.72.094507},
as shown in table \ref{tab:n2hfs}, indicates that the effect of the corrections
is to reduce the lattice spacing dependence to within the remaining errors.
The resulting estimate for the hyperfine splitting of $68(3)(5)(3)$~MeV is
then in good agreement with the experimental value of $69.3(2.8)$~MeV
\cite{PDG:2010}.
Subsequent simulations
\cite{Dowdall:2011wh,Dowdall:2012ab}
have confirmed these expectations. It will be interesting to see how the
inclusion of our radiative corrections in simulations utilizing the $O(v^6)$
NRQCD action will compare with the results of
\cite{Meinel10},
where a reduced hyperfine splitting was found when using the tree-level $O(v^6)$ NRQCD action; 
from our results, we expect that the inclusion of the four-fermion operators will 
compensate for this reduction.

The elimination of $O(\alpha_s a^2)$ errors, the much reduced dependence of observables on
$a^2$ and the agreement with experiment gives us confidence that the improvement strategy
for constructing the NRQCD effective action is robust.

%%%%%%%%%%%%%%%%%%%%%%%%%%%%%%%%%%%%%%%%%%%%%%%%%%%%%%%%%%%%%%%%%%%%%%%%%%%%%%%%
%%%%%%%%%%%%%%%%%%%%%%%%%%%%%%%%%%%%%%%%%%%%%%%%%%%%%%%%%%%%%%%%%%%%%%%%%%%%%%%%
% Acknowledgments
%%%%%%%%%%%%%%%%%%%%%%%%%%%%%%%%%%%%%%%%%%%%%%%%%%%%%%%%%%%%%%%%%%%%%%%%%%%%%%%%
%%%%%%%%%%%%%%%%%%%%%%%%%%%%%%%%%%%%%%%%%%%%%%%%%%%%%%%%%%%%%%%%%%%%%%%%%%%%%%%%
\acknowledgments
We thank Christine Davies, Alan Gray, Andrew Lee, Peter Lepage, John Gracey
and Matthew Wingate for useful discussions.
We thank the DEISA Consortium (\url{http://www.deisa.eu}),
co-funded through the EU~FP6 project RI-031513 and the FP7 project RI-222919,
for support within the DEISA Extreme Computing Initiative. 
This work was supported by STFC under grants ST/G000581/1 and ST/H008861/1.
The calculations for this work were, in part, performed on the University
of Cambridge HPCs as a component of the DiRAC facility jointly funded by
STFC and the Large Facilities Capital Fund of BIS.
The University of Edinburgh is supported in part by the 
Scottish Universities Physics Alliance~(SUPA).

%%%%%%%%%%%%%%%%%%%%%%%%%%%%%%%%%%%%%%%%%%%%%%%%%%%%%%%%%%%%%%%%%%%%%%%%%%%%%%%%
%%%%%%%%%%%%%%%%%%%%%%%%%%%%%%%%%%%%%%%%%%%%%%%%%%%%%%%%%%%%%%%%%%%%%%%%%%%%%%%%
% References
%%%%%%%%%%%%%%%%%%%%%%%%%%%%%%%%%%%%%%%%%%%%%%%%%%%%%%%%%%%%%%%%%%%%%%%%%%%%%%%%
%%%%%%%%%%%%%%%%%%%%%%%%%%%%%%%%%%%%%%%%%%%%%%%%%%%%%%%%%%%%%%%%%%%%%%%%%%%%%%%%
\bibliographystyle{h-physrev4}
\bibliography{bibliography}

\end{document}